\documentclass[oupdraft]{bio}
\usepackage{url}
\usepackage{subcaption}
\usepackage{booktabs}

\history{Received August 1, 2010;
revised October 1, 2010;
accepted for publication November 1, 2010}

\begin{document}

% Title of paper
\title{A Capture-Recapture Approach to Enhance Treatment Effect Evaluation in an Observational Cohort}

% List of authors, with corresponding author marked by asterisk
\author{Lin Ge$^1$, Yuzi Zhang$^2$, Lance A. Waller$^3$, Robert H. Lyles$^3$\\[4pt]
% Author addresses
\textit{$^1$Department of Epidemiology and Biostatistics, School of Public Health,
Indiana University, Bloomington, IN, U.S.A
$^2$Division of Biostatistics, College of Public Health, Ohio State University, Columbus, OH, U.S.A
$^3$Department of Biostatistics and Bioinformatics, Rollins School of Public Health,
Emory University, Atlanta, GA, U.S.A
}
\\[2pt]
% E-mail address for correspondence
{lge\_biostat@outlook.com}}

% Running headers of paper:
\markboth%
% First field is the short list of authors
{L. Ge and others}
% Second field is the short title of the paper
{Capture-Recapture Approach to Enhance Treatment Effect Evaluation}

\maketitle

% Add a footnote for the corresponding author if one has been
% identified in the author list
%\footnotetext{To whom correspondence should be addressed.}

\begin{abstract}
{We extend recently proposed design-based capture-recapture (CRC) methods for prevalence estimation among registry participants, in order to enhance treatment effect evaluation among a trial-eligible target population. The so-called ``anchor stream design" for CRC analysis integrates an observational study cohort with a randomized trial involving a small representative study sample, and enhances the generalizability and transportability of CRC findings. We show that a novel CRC-type estimator derived via multinomial distribution-based maximum-likelihood further exploits the design to deliver benefits in terms of validity and efficiency for comparing the effects of two treatments on a binary outcome. The design also unlocks a direct standardization-type estimator that allows efficient estimation of general means (e.g., for continuous outcomes such as biomarker levels) under a specific treatment. This provides an avenue to compare treatment responses within the target population in a more comprehensive manner. For inference, we recommend using a tailored Bayesian credible interval approach to improve coverage properties in conjunction with the proposed CRC estimator when estimating binary treatment effects, and a bootstrap percentile interval approach for use with continuous outcomes. Simulations demonstrate the validity and efficiency of the proposed estimators under the CRC design. Finally, we present an illustrative data application comparing Anti-S Antibody seropositive response rates for two major Covid-19 vaccines using an observational cohort from Tunisia.}{Capture-Recapture methods, Treatment effect evaluation,
Generalizability and transportability, Standardization}
\end{abstract}

\section{Introduction}
\label{sec1}

Observational studies are widely used in various fields, such as epidemiology and the social sciences, as they facilitate the collection of cohort data for analysis. However, evaluating treatment effects based on observational data is well known to be problematic due to the lack of random experimental assignment to treatments, resulting in confounding bias \citep{Colnet2024}. To reduce bias in observational studies, methods such as propensity matching \citep{DAgostino1998}, inverse probability weighting (IPW) \citep{Robins1994}, or augmented IPW (AIPW) \citep{Robins1994}, are often used in practice. On the other hand, randomized trials offer favorable properties with respect to internal validity \citep{Degtiar2023} and unbiased estimation, although generalizing the conclusions to all eligible individuals can be challenging \citep{Rothwell2005,Dahabreh2019}. Many studies focus on addressing the generalizability from randomized trial results to broader target populations \citep{Hernan2011,Bareinboim2013,Stuart2018}.

In this article, we tailor capture-recapture (CRC) methods toward extending treatment effect evaluation from an observational cohort to a larger registered trial-eligible target population by embedding a relative smaller randomized trial. CRC methodology was originally developed for use in ecological studies seeking to estimate wildlife populations in a specific area \citep{Chao2001,Borchers2002}, but has also been applied in numerous epidemiological and public health research studies for estimating case counts or prevalence of diseases \citep{Wu2005,Dunbar2011,Poorolajal2017} and conditions \citep{Frischer1991}. Key to CRC analysis is estimating the missing count of individuals “caught” by none of the capture efforts, enabling an overall count assessment from the sum of the observed count and the estimated missing count. CRC analysis can be adapted to estimate the mean of a binary outcome (e.g., for prevalence estimation) or a continuous outcome \citep{Lyles2023} (e.g., a biomarker level) in a closed target population, thus potentially making CRC tools useful for addressing treatment effects. 

In this article, the treatment effect evaluation setting is conceptualized within a two-stream CRC design and analysis framework. While sensitivity and uncertainty analyses have been explored \citep{Zhang2020,Zhang2023}, the implementation of CRC analysis based on two data streams is generally problematic without a key independence assumption known as the Lincoln-Petersen, or “LP”, condition \citep{Chao2008}. It assumes that the two data streams utilized in the CRC analysis operate independently of each other, at least at a population level. However, in practice, it is well known that such independence is often questionable and can lead to biased estimation if violated \citep{Seber1982,Brenner1995}. To address this issue, several articles \citep{Seber1986,Chao2008} have discussed how the independence condition can be satisfied by conducting a principled random sample from the target population that is designed to be independent of a second established surveillance effort. Furthermore, the assumption of homogeneous capture probabilities at the individual level in the established data stream can also be relaxed when such an independent random sample is drawn \citep{Chao2008}. One such design-based approach has been proposed recently within CRC analysis \citep{Lyles2022a,Lyles2023,Ge2023} when the target population consists of a list or registry amenable to random sampling, providing a so-called “anchor stream” of representative data that augments a non-representative sample. Because a key association parameter becomes known by design, this approach yields an estimator of population size that is generally far more precise than traditional CRC estimators under the LP conditions \citep{Seber1982}. In this article, we focus on this design-based approach and implement the CRC analysis framework more generally, based on embedding a randomized trial within a larger observational cohort by collecting a relatively small random sample of members of a registered clinic population. Those selected are randomized to one of two available treatments, so that causal conclusions about treatment effects within the target population can be justified based on the presence of the random sample. Because this sample will typically be small for feasibility, however, our goal is to compare treatment success rates in a way that generalizes to the entire target population while also leveraging added precision by including information from arbitrarily non-representative observational data on subjects who utilized the treatment that they or their provider selected.

\begin{figure}
\centering
\begin{subfigure}{.5\textwidth}
  \centering
  \includegraphics[width=1\linewidth]{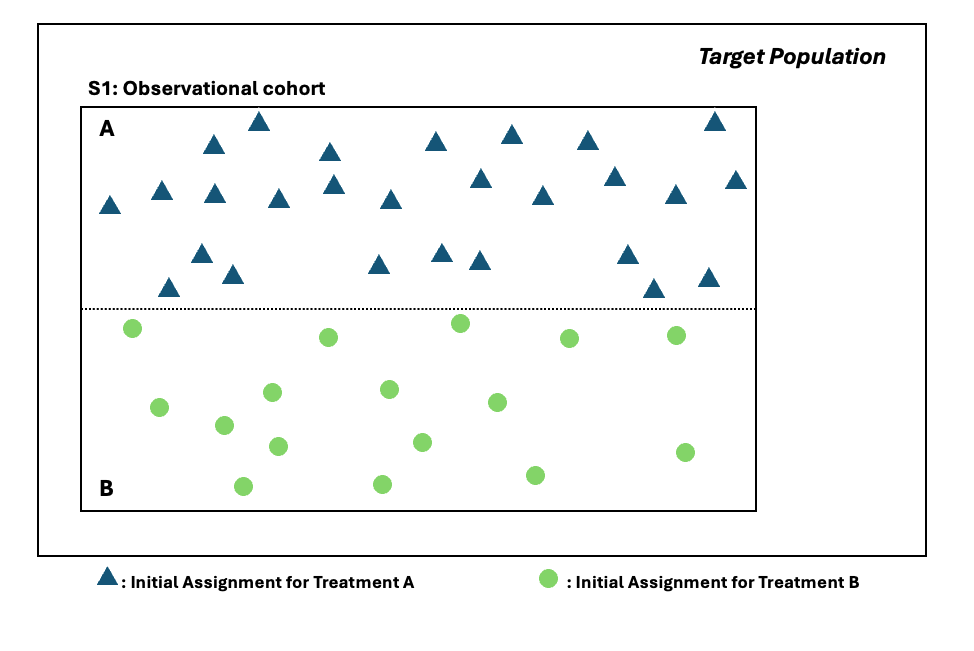}
  \caption*{Fig 1(A)}
  \label{Fig1A}
\end{subfigure}%
\begin{subfigure}{.5\textwidth}
  \centering
  \includegraphics[width=1\linewidth]{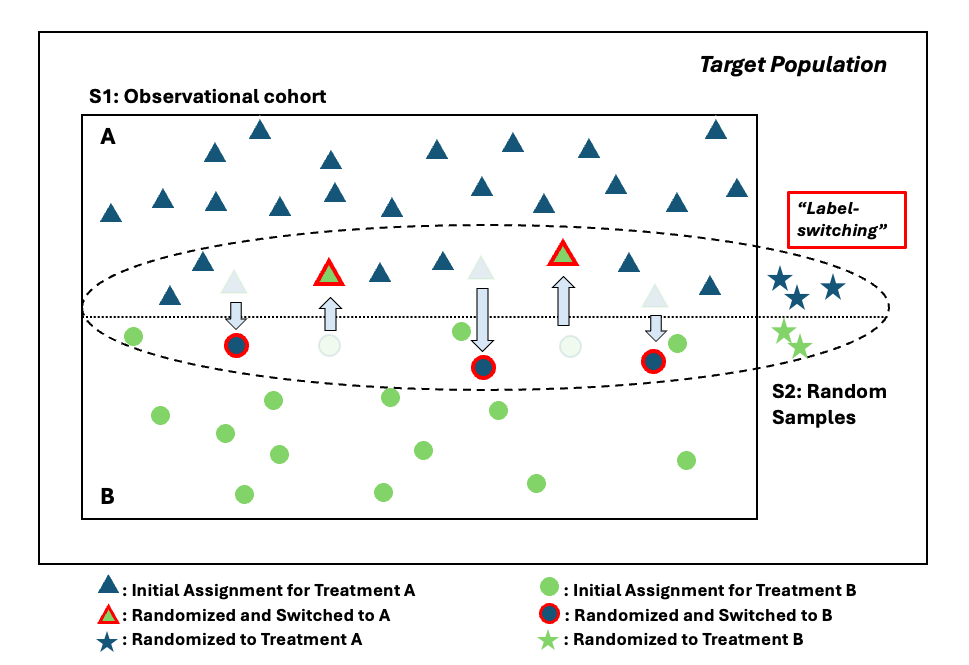}
  \caption*{Fig 1(B)}
  \label{Fig1B}
\end{subfigure}
\caption{Visualization Diagrams Illustrating the Study Design Details Under the “Label-Switching” Strategy. Fig 1(A) shows the initial assigned treatment ($A$ or $B$) in the observational cohort (Stream 1, or S1), as determined by the medical providers. Fig 1(B) indicates the final treatment assignments, incorporating provider “buy-in” for randomly selected individuals from the target population (Stream 2, or S2). The red outlines describe the “label-switching” strategy and highlight those randomized and administered to switch to a different treatment from the initial assigned selection. The star symbols represent new participants with randomized treatment selection by Stream 2 from the target population, who did not receive initial assignment in the observational cohort.}
\label{Fig1}
\end{figure}

The methods that we propose are based on the clinical equipoise assumption \citep{Kukla2007,cook2011} which stipulates that there is no established preference for one treatment over another in a given population. This setting is common in trial design (e.g., when studying the repurposing of approved drugs) and has been leveraged, for example, to compare option $A$ and option $B$ drug regimens for prevention of mother-to-child transmission of HIV (PMTCT) \citep{sando2014}. We assume that two treatments in equipoise are being evaluated among a closed target population of individuals eligible for both treatments. An observational study is to be initiated (forming the basis for the first data stream), and a small randomized trial is essentially designed to be embedded within the population in order to obtain treatment-specific outcomes for a representative subset of participants selected from the target population \citep{Shadish2002,Olsen2016} using either a simple or stratified random sampling approach. The CRC method combines information from both the observational and experimental data, achieving dual goals: improving the reliability of the observational evidence via the randomized trial data, and increasing the statistical efficiency of the randomized trial component via the observational information \citep{Colnet2024}. This study design is detailed as follows, and visualized in Figure \ref{Fig1}.

\begin{itemize}
    \item \textit{We assume a closed trial-eligible participant population with a known size, within which both equipoised treatment options ($A$ and $B$) are to be made available to participants. Individuals who could not feasibly be administered one or both treatments of interest (e.g., due to indications associated with risk or tolerability) are first removed from the target population.}
    \item \textit{The medical providers of the participants determine an assignment of treatment ($A$ or $B$). This assignment may be associated with physician preference (possibly driven by ties to the manufacturer) and driven by participant characteristics (e.g., clinical data, insurance status, etc.) that could be related to the probability of treatment response. This forms the basis of the observational cohort subject to initial assigned treatment selection by the provider, which we hereafter refer to as Stream 1 (or S1). Note that the observational cohort is a subset of the target population in the study.}
    \item \textit{Before initiating the assigned treatment ($A$ or $B$) for each person by the provider, we collect a random sample from the target population. In size, this sample will typically be small relative to the observational cohort. Each selected participant is then randomized to either receive $A$ or $B$, forming the basis for the sampling-based component that we denote as Stream 2 (or S2). Importantly, we assume the buy-in of the observational cohort and their providers. That is, if a patient is selected in the random sample and randomized to the treatment not initially chosen, their provider will administer the randomly prescribed treatment. This is referred to as a “label-switching” strategy in the following sections; however, it is important to note that all other patients will keep the treatment initially assigned by the provider.}
\end{itemize}

\section{Methods} \label{sec2}
\subsection{Notations and Assumptions} \label{sec2.1}

Let \(\mathcal{T}\) be the set of treatments of interest to be assessed in the CRC analysis. For simplicity, we only consider two treatments (\(A\) and \(B\)) here; however, an extension to more treatments can be made naturally. We use the following notation to facilitate the description of the proposed method: \(N_{tot}\) is the known total size of the closed target population, and \(i = 1,2,\ldots,\ N_{tot}\) indexes each individual in the population; \(S_{i}^{(1)}\), \(S_{i}^{(2)}\) are the indicators for being observed in the observational cohort or the anchor stream respectively; \(T_{i}^{(1)}\) is the treatment assignment in the observational cohort study, \(T_{i}^{(2)}\) is the randomized treatment assignment in the anchor stream, \(T_{i}\) is the final treatment assignment after the \emph{``label-switching''} strategy (i.e., \(T_{i} = T_{i}^{(2)}\) if the patient has an assigned treatment \(T_{i}^{(2)}\) from the anchor stream; otherwise \(T_{i} = T_{i}^{(1)}\)). Note that \(T_{i}\) can be a missing value if neither of the two streams assigns a treatment. \(Y_{i}\) is the observed binary outcome to indicate treatment response and \({Y_{i}}^{A}\) is the potential outcome for each \(A \in \mathcal{T}\); this is extended to the case of continuous outcome treatment response evaluation \({{\widetilde{Y}}_{i}}^{A}\) in Section \ref{sec2.5}.

\begin{figure}[t]
    \centering
    \includegraphics[width=0.8\linewidth]{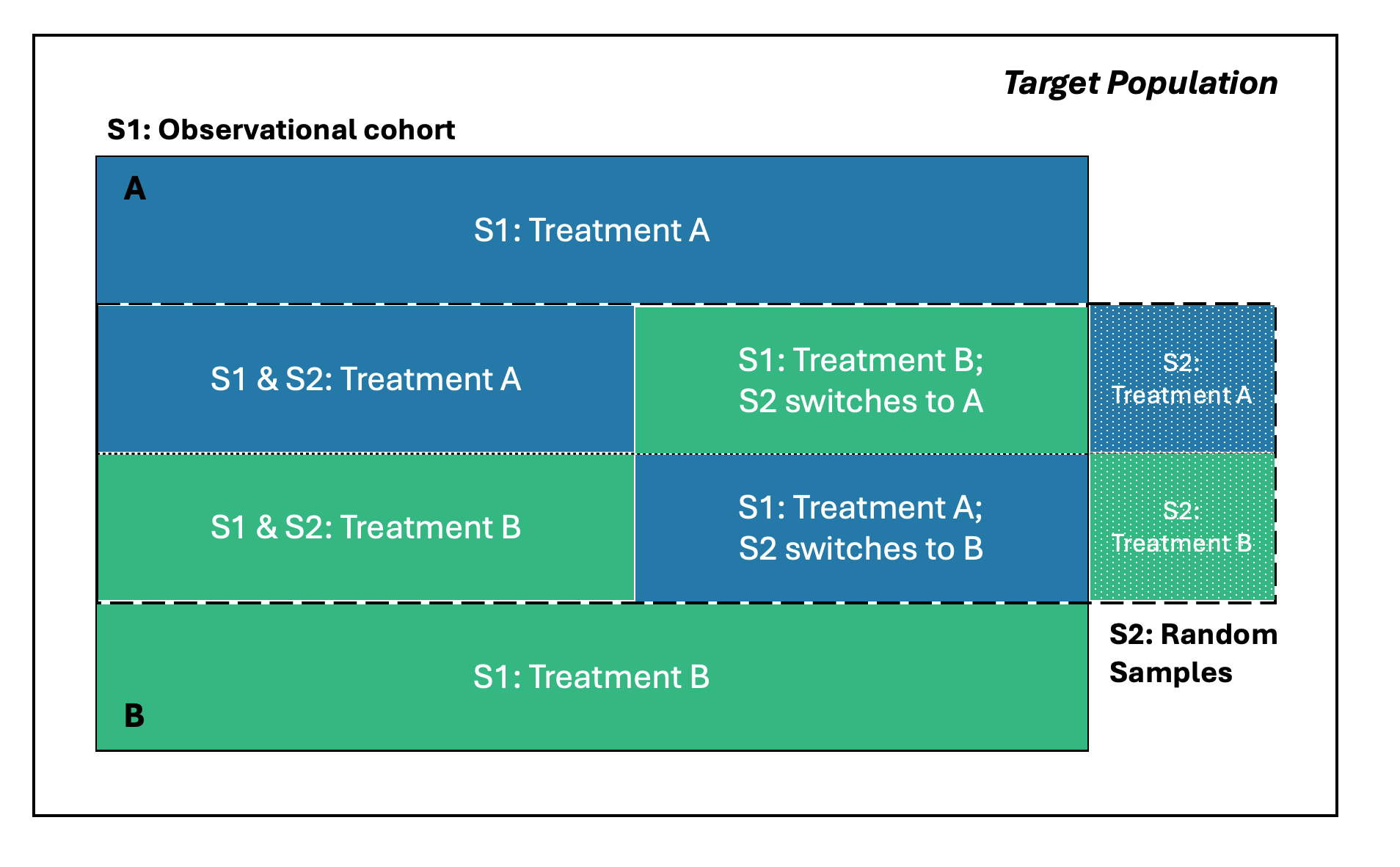}
    \caption{Visualization Diagram illustrating the CRC observations based on the study design.}
    \label{Fig2}
\end{figure}

We are interested in evaluating treatment effects with respect to the entire target population based on the observed CRC data from Streams 1 (observational cohort) and 2 (anchor stream). To better visualize the data from this study design, we use Figure \ref{Fig2} to illustrate the CRC observations. In general, the targets of estimation are the potential outcome means \(\mu(A) = E(Y^{A})\) for each \(A \in \mathcal{T}\), and the Average Treatment Effects (\(ATE\)), i.e., \(ATE = \ E(Y^{A}) - E(Y^{B})\) for any pair of treatments \(A,\ B \in \mathcal{T}\).

In what follows, we assume the design strategy described above to be in effect. Implementing this design implies satisfying the assumptions such as treatment  positivity in Stream 1 and external validity with sampling ignorability in Stream 2. At the same time, it enables us to relax several crucial assumptions typically required for drawing causal conclusions in Stream 1, such as conditional ignorability \citep{Hernan2020,Parikh2024}. Nevertheless, the consistency assumption remains essential \citep{Hernan2020}, namely, that the potential outcome under a specific treatment is equal to the observed outcome when receiving that treatment. While valid inference can be achieved using the representative Stream 2 sample alone under these assumptions, a key objective is to also leverage the likely much larger but arbitrarily non-representative Stream 1 sample in the interest of improved precision.

\subsection{Existing Classical Estimators for Treatment Effect Estimation}\label{sec2.2}

The ``anchor stream'' design with label switching ensures that the
sample subject to randomization in Stream 2 is drawn ``agnostically''
(i.e., independently) with respect to the ultimate Stream 1
observational study participant cohort subset of the target population. As Stream 2 introduces representative samples of the target population assigned to each treatment, a simple and defensible estimand is immediately available to satisfy the identifiability of the potential outcome mean \(\mu(A)\), i.e., \(E\left( Y \middle| T = A,S^{(2)} = 1 \right)\) for each \(A \in \mathcal{T}\). The random sampling-based estimator is given as follows:
\begin{align}\label{eq.1}
    {\hat{\mu}}_{RS}(A) = \frac{1}{n_{A}}\sum_{i}^{}{I(T_{i}^{(2)} = A)Y_{i}},\ \ \hat{V}\left( {\hat{\mu}}_{RS}(A) \right) = \frac{{\hat{\mu}}_{RS}(A)\ \left( 1 - {\hat{\mu}}_{RS}(A) \right)}{n_{A}}
\end{align}
where \(n_{A} = \sum_{i}^{}{I(T_{i}^{(2)} = A)}\) and \(A \in \mathcal{T}\). Note that in this study setting, we assume no finite population correction (FPC) to the variance in (\ref{eq.1}), in contrast to previous studies \citep{Lyles2022a,Lyles2023,Ge2023,Ge2024} of anchor stream-based disease prevalence estimation that incorporated FPC effects. This aligns with the implications of sampling repetition in the current setting, where the total population and the S1 and S2 sample sizes remain fixed but the number of individuals responding to treatment
varies.

Alternatively, note that implementation of the anchor stream design fully justifies the Lincoln-Petersen condition \citep{Lincoln1930,Petersen1986}, so that the classical LP and Chapman CRC estimators are applicable \citep{Lincoln1930,Chapman1951,Petersen1986}. One focus of this article is on estimating potential outcome means for a given treatment \(A \in \mathcal{T}\) among the target population, which is equivalent to estimating the response rate from the estimated total outcomes of responding individuals. This could be done via \((\ref{eq.1})\) or via simple CRC estimation based on the three observed cell counts in Table \ref{tbl.1} \citep{Agresti1998}. Specifically for treatment \(A \in \mathcal{T}\), \(n_{11}\) is the number of responders among those who finally receive treatment \(A\) in both Stream 1 and Stream 2 (i.e., \(T = A\), \(S^{(1)} = 1\), and \(S^{(2)} = 1\)). The cell count \(n_{10}\) is the number of responders among those who finally receive treatment \(A\) in Stream 1 but were not selected for randomization in Stream 2 (i.e., \(T = A\), \(S^{(1)} = 1\), and \(S^{(2)} = 0\)). Lastly, \(n_{01}\) is the number of responders among those who were not finally assigned to treatment \(A\) in Stream 1 but were selected for randomization to treatment \(A\) in Stream 2 (i.e., \(T = A\), \(S^{(1)} = 0\), and \(S^{(2)} = 1\)). Note that the potential outcome means can be approximated by the response rate averaging across the known effective total population size \(N_{tot}^{(A)}\) of the target population \(N_{tot}\), which is equal to \(N_{tot}\) minus the number of individuals selected and randomized to treatment \(B\) in Stream 2. That is, the anchor stream sampling procedure wherein individuals are first randomly chosen for Stream 2 and subsequently randomized for both treatments, is equivalent to a process in which individuals are randomly chosen for treatment \(A\) from a subpopulation that omits those randomly chosen for treatment \(B\) in Stream 2.

It follows that one direct and valid alternative to (\ref{eq.1}) for estimating the potential outcome mean for treatment \(A\) is to use the well-known Chapman estimator \citep{Chapman1951,Seber1986}, i.e., \({\hat{\mu}}_{Chap}(A)\) and its estimated variance are given as follows:
\begin{align}\label{eq.2}
    {\hat{\mu}}_{Chap}(A) = \frac{1}{N_{tot}^{(A)}}\left\lbrack \frac{{(n}_{1.} + 1)(n_{.1} + 1)}{n_{11} + 1} - 1 \right\rbrack,\ \ \hat{V}\left( {\hat{\mu}}_{Chap}(A) \right) = \frac{1}{{N_{tot}^{(A)}}^{2}}\left\lbrack \frac{{(n}_{1.} + 1)(n_{.1} + 1)n_{1.}n_{.1}}{(n_{11} + 1)^{2}(n_{11} + 2)} \right\rbrack\ 
\end{align}

Wald-type confidence intervals (CIs) based on the Chapman estimate and its variance in \((\ref{eq.2})\) are known to provide unsatisfactory coverage in many CRC settings under the LP conditions. Therefore, we summarize results based on a more reliable transformed logit CI \citep{Sadinle2009} in our simulation studies to follow.

\begin{table}
\centering
\caption{Responder Counts for Two-Stream Capture-Recapture for One Treatment $T = A \in \mathcal{T}$}\label{tbl.1}
\renewcommand{\arraystretch}{1.1}
\setlength{\tabcolsep}{5pt}
\begin{tabular}{ccccc}
\toprule
\multirow{2}{*}{ }
& \multicolumn{2}{c}{Observed to respond to treatment $A$ in }  &\\
&  \multicolumn{2}{c}{Stream 2 (i.e., $S^{(2)}=1$)} &\\
\cmidrule(lr){2-3}
Observed to respond to treatment & \multirow{2}{*}{Yes} & \multirow{2}{*}{No} & \multirow{2}{*}{Total} & \\
$A$ in Stream 1 (i.e., $S^{(1)}=1$) &&& \\
\midrule
Yes   & $n_{11}$ & $n_{10}$ & $n_{1\cdot}$ & \\ 
No    & $n_{01}$ & $n_{00} \,=?$ & & \\ 
\midrule
Total & $n_{\cdot 1}$ & & $N \,=?$ \\ 
\bottomrule
\end{tabular}
\vspace{0.6em}

\raggedright
\footnotesize \textit{* The effective population size is $N^{(A)}_{\text{tot}}$, which omits individuals chosen and randomized for another treatment by Stream 2 from the target population ($N_{\text{tot}}$).}
\end{table}

\subsection{More Efficient CRC Estimators for Treatment Effect Estimation}\label{sec2.3}

An alternative CRC estimator can be developed by utilizing the full
observation profile obtained by design in the anchor stream CRC setting with the \emph{``label-switching''} strategy (see Figure \ref{Fig2}).
Specifically, the design yields a maximum-likelihood estimator (MLE) for the potential outcome means of both treatments simultaneously, i.e.,
\(\mu(A)\) and \(\mu(B)\), under a 17-category multinomial distribution
that accounts for each of the \(N_{tot}\) members of the registered
target population. The details of each observed cell count and its
likelihood contribution are in Table \ref{tbl.2}. The derivation for each likelihood contribution is available in Appendix 1 of
{\textit{Supplementary Materials}}.

\begin{table}[ht]
    \centering
    \caption{Cell Counts and Likelihood Contributions for Observations}
    \label{tbl.2}
    \begin{tabular}{lll}
    \toprule
    Cell & \multirow{2}{*}{Observation Type } &  Multinomial likelihood  \\
     Count &    &  contribution\\
     \midrule
    \multirow{2}{*}{$n_1$} & Sampled in both streams, assigned and randomized to & \multirow{2}{*}{\(p_{1} = \xi_{A}\psi\pi_{s1,A}\phi_{A}\phi\)} \\
    & \(A\), \(Y\  = \ 1\) & \\
    
    \multirow{2}{*}{$n_2$} & Sampled in both streams, assigned and randomized to & \multirow{2}{*}{\(p_{2} = \xi_{A}\psi(1 - \pi_{s1,A})\phi_{A}\phi\)} \\
    & \(A\), \(Y\  = \ 0\) & \\

    \multirow{2}{*}{$n_3$} & Sampled and assigned treatment \(A\) in Stream 1, but not & \multirow{2}{*}{\(p_{3} = (1 - \psi)\pi_{s1,A}\phi_{A}\phi\)} \\
    & sampled in Stream 2, \(Y\  = \ 1\) & \\

    \multirow{2}{*}{$n_4$} & Sampled and assigned treatment \(A\) in Stream 1, but not & \multirow{2}{*}{$p_{4}=(1-\psi)(1-\pi_{S1,A})\phi_A\phi$} \\
    & sampled in Stream 2, \(Y\  = \ 0\) & \\

    \multirow{2}{*}{$n_5$} & Sampled and assigned treatment $B$ in Stream 1; sampled & \multirow{2}{*}{$p_{5}=\xi_A\psi\pi_{S1,B,A}(1-\phi_A)\phi$} \\
    & in Stream 2, randomized and switched label to $A$, $Y=1$ & \\  

    \multirow{2}{*}{$n_6$} & Sampled and assigned treatment $B$ in Stream 1; sampled & \multirow{2}{*}{$p_{6}=\xi_A\psi(1-\pi_{S1,B,A})(1-\phi_A)\phi$} \\
    & in Stream 2, randomized and switched label to $A$, $Y=0$ & \\

    \multirow{2}{*}{$n_7$} & Sampled in both streams, assigned and randomized to & \multirow{2}{*}{$p_{7}=(1-\xi_A)\psi\pi_{S1,B}(1-\phi_A)\phi$} \\
    & $B$, $Y=1$ & \\

    \multirow{2}{*}{$n_8$} & Sampled in both streams, assigned and randomized to & \multirow{2}{*}{$p_{8}=(1-\xi_A)\psi(1-\pi_{S1,B})(1-\phi_A)\phi$} \\
    & $B$, $Y=0$ & \\

    \multirow{2}{*}{$n_9$} & Sampled and assigned treatment $B$ in Stream 1, but not & \multirow{2}{*}{$p_{9}=(1-\psi)\pi_{S1,B}(1-\phi_A)\phi$} \\
    & sampled in Stream 2, $Y=1$ & \\

    \multirow{2}{*}{$n_{10}$} & Sampled and assigned treatment $B$ in Stream 1, but not & \multirow{2}{*}{$p_{10}=(1-\psi)(1-\pi_{S1,B})(1-\phi_A)\phi$} \\
    & sampled in Stream 2, $Y=0$ & \\

    \multirow{2}{*}{$n_{11}$} & Sampled and assigned treatment $A$ in Stream 1; sampled & \multirow{2}{*}{$p_{11}=(1-\xi_A)\psi\pi_{S1,A,B}\phi_A\phi$} \\
    & in Stream 2, randomized and switched label to $B$, $Y=1$ & \\

    \multirow{2}{*}{$n_{12}$} & Sampled and assigned treatment $A$ in Stream 1; sampled & \multirow{2}{*}{$p_{12}=(1-\xi_A)\psi(1-\pi_{S1,A,B})\phi_A\phi$} \\
    & in Stream 2, randomized and switched label to $B$, $Y=0$ & \\

    \multirow{2}{*}{$n_{13}$} & Not sampled in Stream 1; sampled in Stream 2 and & \multirow{2}{*}{$p_{13}=\xi_A\psi\pi_{S1,NA,A}(1-\phi)$ } \\
    & randomized to $A$, $Y=1$ & \\

    \multirow{2}{*}{$n_{14}$} & Not sampled in Stream 1; sampled in Stream 2 and & \multirow{2}{*}{$p_{14}=\xi_A\psi(1-\pi_{S1,NA,A})(1-\phi)$ } \\
    & randomized to $A$, $Y=0$ & \\

    \multirow{2}{*}{$n_{15}$} & Not sampled in Stream 1; sampled in Stream 2 and & \multirow{2}{*}{$p_{15}=(1-\xi_A)\psi\pi_{S1,NA,B}(1-\phi)$ } \\
    & randomized to $B$, $Y=1$ & \\

    \multirow{2}{*}{$n_{16}$} & Not sampled in Stream 1; sampled in Stream 2 and & \multirow{2}{*}{$p_{16}=(1-\xi_A)\psi(1-\pi_{S1,NA,B})(1-\phi)$ } \\
    & randomized to $B$, $Y=0$ & \\

    $n_{17}$ & Not sampled in Stream 1; not sampled in Stream 2 & $p_{17}=(1-\psi)(1-\phi)$ \\
    \bottomrule
    \end{tabular}
\vspace{0.6em}

\raggedright
\footnotesize \textit{* Connections with Table \ref{tbl.1}:
\(n_{11}^{(A)} = n_{1}\), \(n_{10}^{(A)} = n_{3}\), and
\(n_{01}^{(A)} = n_{5} + n_{13}\); \(n_{11}^{(B)} = n_{7}\),
\(n_{10}^{(B)} = n_{9}\), and \(n_{01}^{(B)} = n_{11} + n_{15}\)}
\end{table}

The likelihood contributions given in Table \ref{tbl.2} are based on defining the following parameters. 
\begin{itemize}
    \item \(\phi\) = Pr(Sampled in Stream 1)
    \item \(\phi_{A}\) = Pr(Assigned treatment \(A\) \textbar{} sampled in Stream 1)
    \item \(\ \pi_{s1,A}\) = Pr($Y=1$ \textbar{} Sampled and assigned treatment \(A\) in Stream 1, and received \(A\))
    \item \(\pi_{s1\_ B,A}\) = Pr($Y=1$ \textbar{} Sampled and assigned treatment \(B\) in Stream 1, but received \(A\))
    \item \(\pi_{s1\_NA,A}\) = Pr($Y=1$ \textbar{} Not sampled in Stream 1, but received \(A\))
    \item \(\pi_{s1,B}\) = Pr(\(Y\  = \ 1\ \)\textbar{} Sampled and assigned treatment \(B\) in Stream 1, and received \(B\))
    \item \(\pi_{s1\_ A,B}\) = Pr($Y=1$ \textbar{} Sampled and assigned treatment \(A\) in Stream 1, but received \(B\))
    \item \(\pi_{s1\_ NA,B}\) = Pr($Y=1$ \textbar{} Not sampled in Stream 1, but received \(B\))
\end{itemize}

%\(\phi\) = Pr(Sampled in Stream 1), \(\phi_{A}\) = Pr(Assigned treatment \(A\) \textbar{} sampled in Stream 1),\(\ \pi_{s1,A}\) = Pr($Y=1$ \textbar{} Sampled and assigned treatment \(A\) in Stream 1, and received \(A\)), \(\pi_{s1\_ B,A}\) = Pr($Y=1$ \textbar{} Sampled and assigned treatment \(B\) in Stream 1, but received \(A\)), \(\pi_{s1\_ NA,A}\) = Pr($Y=1$ \textbar{} Not sampled in Stream 1, but received \(A\)), \(\pi_{s1,B}\) = Pr(\(Y\  = \ 1\ \)\textbar{} Sampled and assigned treatment \(B\) in Stream 1, and received \(B\)), \(\pi_{s1\_ A,B}\) = Pr($Y=1$ \textbar{} Sampled and assigned treatment \(A\) in Stream 1, but received \(B\)), \(\pi_{s1\_ NA,B}\) = Pr($Y=1$ \textbar{} Not sampled in Stream 1, but received \(B\)). 
Additionally, there are two more parameters \(\psi\) = Pr(Sampled in Stream 2) and \(\xi_{A}\) = Pr( Randomized to \(A\) \textbar{} Sampled in Stream 2) that can be treated as known. We set \(\xi_{A} = 50\%\) here to reflect an assumption of balanced treatment assignment via randomization, but it can be altered to accommodate unbalanced scenarios targeted by design. Letting \(p_{j}\) denote the likelihood contribution corresponding to the \(j\)th cell, the vector of cell counts can be modeled as a multinomial sample with likelihood proportional to \(\prod_{j = 1}^{17}p_{j}\), i.e.,
\begin{align*}
    \left( n_{1},\ n_{2},\ \cdots,n_{17} \right)\sim multinomial\left( N_{tot};\ p_{1},p_{2},\cdots,p_{17} \right)
\end{align*}
All parameters in Table \ref{tbl.2} are identifiable, and the MLEs of each parameter as well as the corresponding estimated variances are derivable in closed form (see Appendix 2) as follows:
\begin{itemize}
\item
  \(\hat{\phi} = \frac{N_{1}}{N_{tot}}\),
  \(\hat{V}( \hat{\phi} ) = \frac{\hat{\phi}\left( 1 - \hat{\phi} \right)}{N_{tot}}\),
  where
  \(N_{1} = N_{tot} - (n_{13} + n_{14} + n_{15} + n_{16} + n_{17})\)
\item
  \({\hat{\phi}}_{A} = \frac{N_{1,A}}{N_{1}}\),
  \(\hat{V}( {\hat{\phi}}_{A} ) = \frac{{\hat{\phi}}_{A}\left( 1 - {\hat{\phi}}_{A} \right)}{N_{1}}\),
  where \(N_{1,A} = n_{1} + n_{2} + n_{3} + n_{4}{+ n}_{11} + n_{12}\)
\item
  \({\hat{\pi}}_{s1,A} = \frac{n_{1} + n_{3}}{n_{1} + n_{2} + n_{3} + n_{4}}\),
  \(\hat{V}\left( {\hat{\pi}}_{s1,A} \right) = \frac{{\hat{\pi}}_{s1,A}\left( 1 - {\hat{\pi}}_{s1,A} \right)}{n_{1} + n_{2} + n_{3} + n_{4}}\)
\item
  \({\hat{\pi}}_{s1\_ B,A} = \frac{n_{5}}{n_{5} + n_{6}}\),
  \(\hat{V}\left( {\hat{\pi}}_{s1\_ B,A} \right) = \frac{{\hat{\pi}}_{s1\_ B,A}(1 - {\hat{\pi}}_{s1\_ B,A})}{n_{5} + n_{6}}\)
\item
  \({\hat{\pi}}_{s1\_ NA,A} = \frac{n_{13}}{n_{13} + n_{14}}\),
  \(\hat{V}\left( {\hat{\pi}}_{s1\_ NA,A} \right) = \frac{{\hat{\pi}}_{s1\_ NA,A}\left( 1 - {\hat{\pi}}_{s1\_ NA,A} \right)}{n_{13} + n_{14}}\)
\item
  \({\hat{\pi}}_{s1,B} = \frac{n_{7} + n_{9}}{n_{7} + n_{8} + n_{9} + n_{10}}\),
  \(\hat{V}\left( {\hat{\pi}}_{s1,B} \right) = \frac{{\hat{\pi}}_{s1,B}\left( 1 - {\hat{\pi}}_{s1,B} \right)}{n_{7} + n_{8} + n_{9} + n_{10}}\)
\item
  \({\hat{\pi}}_{s1\_ A,B} = \frac{n_{11}}{n_{11} + n_{12}}\),
  \(\hat{V}\left( {\hat{\pi}}_{s1\_ A,B} \right) = \frac{{\hat{\pi}}_{s1\_ A,B}(1 - {\hat{\pi}}_{s1\_ A,B})}{n_{11} + n_{12}}\)
\item
  \({\hat{\pi}}_{s1\_ NA,B} = \frac{n_{15}}{n_{15} + n_{16}}\),
  \(\hat{V}\left( {\hat{\pi}}_{s1\_ NA,B} \right) = \frac{{\hat{\pi}}_{s1\_ NA,B}\left( 1 - {\hat{\pi}}_{s1\_ NA,B} \right)}{n_{15} + n_{16}}\)
\end{itemize}
Of special note and convenience here is the fact that the covariances among the 8 closed-form MLEs above are all equal to zero under the multinomial model.

The estimated potential outcome means under each treatment option (generalizable to the entire target population) are evaluated as
follows:
\begin{align}
    \ {\hat{\mu}}_{CRC}(A) &= {\hat{\pi}}_{s1,A}{\hat{\phi}}_{A}\hat{\phi} + {\hat{\pi}}_{s1\_ B,A}\ (1 - {\hat{\phi}}_{A})\hat{\phi} + {\hat{\pi}}_{s1\_ NA,A}\ (1 - \hat{\phi}) \label{eq.3} \\
    {\hat{\mu}}_{CRC}(B) &= {\hat{\pi}}_{s1,B}\left( 1 - {\hat{\phi}}_{A} \right)\hat{\phi} + {\hat{\pi}}_{s1\_ A,B}\ {\hat{\phi}}_{A}\hat{\phi} + {\hat{\pi}}_{s1\_ NA,B}\ (1 - \hat{\phi}) \label{eq.4}
\end{align}

The variance estimators for (\ref{eq.3}) and (\ref{eq.4}) are readily evaluated via the multivariate delta method, facilitated by available closed forms for the variances of the individual estimated parameters. Details about the derivations are available in Appendix 3 of the {\textit{Supplementary Materials}}. To achieve better coverage rates for interval estimation, we propose a Bayesian credible interval approach in the following section to improve upon the ordinary Wald-type confidence interval.

One advantage of this CRC estimator based on the multinomial distribution underlying Table \ref{tbl.2} is that both of the treatment effects can be estimated simultaneously. To connect directly with past work on use of the anchor stream design for prevalence estimation, one can also derive maximum likelihood estimators corresponding to each treatment based on two separate condensed versions of Table \ref{tbl.2}. This involves collapsing the observation profiles in terms of a single treatment and denoting those who were randomized to the other treatment as ``not sampled in either stream'', as in Table 2 of \cite{Lyles2022a}. This yields alternative estimates of the
potential outcome means for both treatments as follows:
\begin{align}
    {\hat{\mu}}_{\hat{\Psi}}(A) &= \frac{1}{N_{tot}^{(A)}}\left\lbrack n_{11,A} + n_{10,A} + \frac{n_{01,A}}{\hat{\Psi}(A)} \right\rbrack \label{eq.5}\\
    {\hat{\mu}}_{\hat{\Psi}}(B) &= \frac{1}{N_{tot}^{(B)}}\left\lbrack n_{11,B} + n_{10,B} + \frac{n_{01,B}}{\hat{\Psi}(B)} \right\rbrack \label{eq.6}
\end{align}
where \(\hat{\Psi}(A) = \frac{n_{5} + n_{6} + n_{13} + n_{14}}{N_{tot}^{(A)} - (n_{1} + n_{2} + n_{3} + n_{4})}\), \(\hat{\Psi}(B) = \frac{n_{11} + n_{12} + n_{15} + n_{16}}{N_{tot}^{(B)} - (n_{7} + n_{8} + n_{9} + n_{10})}\), \(N_{tot}^{(A)} = N_{tot} - (n_{7} + n_{8} + n_{11} + n_{12} + n_{15} + n_{16})\) and \(N_{tot}^{(B)} = N_{tot} - (n_{1} + n_{2} + n_{5} + n_{6} + n_{13} + n_{14})\). Note that the connection between the observed cell counts in
Table \ref{tbl.1} and Table \ref{tbl.2} is given in the footnote of Table \ref{tbl.2}. The accompanying variance estimators for (\ref{eq.5}) and (\ref{eq.6}) are evaluated through the multivariate delta method based on the condensed table specific to each single treatment \citep{Lyles2022a}.

\subsection{A Bayesian Credible Interval Approach}\label{sec2.4}

The performance of Wald-type confidence intervals (CIs) in binomial/multinomial settings has been shown to be unsatisfactory in numerous studies \citep{Agresti1998,Brown2001}, especially when the sample size is small. In this article, we propose a Bayesian credible interval approach based on a weakly informative Jeffreys prior on the full multinomial model associated with Table \ref{tbl.2} in an effort to provide more reliable coverage compared to Wald-type CIs as companions to the novel multinomial distribution-based CRC estimator introduced above. The approach has connections with similar proposals made in conjunction with the original anchor stream design for estimating a prevalent case count \citep{Lyles2022a,Lyles2023}, except in this case the administration of treatment followed by outcome assessment allows one to rely upon the typical multinomial variance-covariance matrix without concern about finite population sampling.

Our proposed credible interval approach begins with a conjugate Jeffreys $Dirichlet$(0.5, 0.5, $\cdots$, 0.5) prior for the 17 cell probabilities associated with Table \ref{tbl.2}, yielding the following
posterior:
\begin{align}
    \left( p_{1}^{*},\ p_{2}^{*},\cdots,p_{17}^{*} \right)|N_{tot}\sim Dirichlet\left( n_{1} + 0.5,\ n_{2} + 0.5,\ \cdots,n_{17} + 0.5 \right) \label{eq.7}
\end{align}

From each posterior draw via (\ref{eq.7}), we derive posterior cell counts \(\left( n_{1}^{*},\ n_{2}^{*},\cdots,n_{17}^{*} \right)\) by multiplying by \(N_{tot}\). Thereafter, we evaluate a posterior draw of the ML estimate \({\hat{\mu}}_{CRC}^{*}\) for each treatment, by inserting the 17 posterior cell counts into (\ref{eq.3}) and (\ref{eq.4}). Similarly, the alternative ML estimator \({\hat{\mu}}_{\hat{\Psi}}^{*}\) based on the separate condensed versions of Table \ref{tbl.2} can be mimicked based on (\ref{eq.5}) and (\ref{eq.6}). Subsequently, the proposed approach reports a (2.5$th$, 97.5$th$) percentile interval based on the posteriors of the estimated treatment effects, as the Bayesian credible interval to accompany each estimator.

\subsection{Extension to Estimate General Means of Continuous Treatment Outcomes} \label{sec2.5}

Now we extend our interest to estimating a treatment effect characterized in terms of the mean of a continuous outcome \(Y\), \(E({\widetilde{Y}}^{A})\), under intervention via treatment \(A\). For example, the treatment outcome might be a continuous variable (e.g., a continuous biomarker level or the change in such a level). In general, the direct standardization method \citep{Naing2000} is a useful approach to estimate means or rates based on stratified sampling with known or estimable sampling rates within strata. Due to the anchor stream design, stratification can be based on three parts that are same
as in the partition represented in (\ref{eq.3}) and (\ref{eq.4}). In this case, a tailored direct standardization-type estimator \citep{Lyles2023} is unlocked for the potential mean of \({\widetilde{Y}}^{A}\) for each
treatment \(A \in \mathcal{T}\) as follows:
\begin{align}
    E\left( {\widetilde{Y}}^{A} \right) = {\overline{y}}_{s1,A}{\hat{p}}_{s1,A} + {\overline{y}}_{s1\_ B,A}{\hat{p}}_{s1\_ B,A} + {\overline{y}}_{s1\_ NA,A}{\hat{p}}_{s1\_ NA,A} \label{eq.8}
\end{align}
where \({\hat{p}}_{s1,A} = {\hat{\phi}}_{A}\hat{\phi}\), \({\hat{p}}_{s1\_ B,A} = (1 - {\hat{\phi}}_{A})\hat{\phi}\), \({\hat{p}}_{s1\_ NA,A} = (1 - \hat{\phi})\) and \({\overline{y}}_{s1,A}\)= E(\(\widetilde{Y}\) \textbar{} Sampled and assigned treatment \(A\) in Stream 1, and received \(A\)), \({\overline{y}}_{s1\_ B,A}\) = E(\(\widetilde{Y}\) \textbar{} Sampled and assigned treatment \(B\) in Stream 1, but received \(A\)), \({\overline{y}}_{s1\_ NA,A}\) = E(\(\widetilde{Y}\) \textbar{} Not sampled in Stream 1, but received \(A\)). The estimators \({\hat{\phi}}_{A}\) and \(\hat{\phi}\) are evaluated based on the MLEs provided in Section \ref{sec2.3} and the expectations are estimated based on sample means of each subpopulation.

Regarding inference on the general mean of a continuous outcome, we propose employing a standard bootstrap approach \citep{Efron1994} on the observed data to assess both the standard error (SE) and the bootstrap percentile intervals. To elaborate, we initiated the process with the observed data records of all individuals identified at least once from either the observational cohort or the anchor stream (i.e., \(S_{i}^{(1)} = 1\) and/or \(S_{i}^{(2)} = 1\) for all \(i\) in the target population). We then randomly draw \(M\) bootstrap samples with replacement from this list of individuals. For each bootstrap sample, we evaluate the estimator using (\ref{eq.8}) and subsequently calculate the standard error and 95\% percentile interval.

\section{Simulation Studies}\label{sec3}

In this section, we present two sets of simulation studies to compare the performance of each estimator introduced in the previous section, assessing the binary treatment effect estimators and the proposed general mean estimator in the case of continuous treatment outcomes.

With sampling under the anchor stream design, we generate the data based on a hypothetical scenario that mimics a comparison between experimental treatments \(A\) and \(B\) among a closed target population. First, we generate a population of size \(N_{tot} = (500;\ 1,000;\ 5,000)\) and stratify it into two groups (40\% vs 60\%) based on a binary characteristic or trait. For Stream 1, we randomly include 70\% of individuals from stratum 1 and 90\% from stratum 2 to form the observational cohort. We initially simulate the assigned treatment by the provider based on the existing strata categories. Specifically, 30\% of individuals in stratum 1 choose treatment \(A\) and 80\% of individuals in stratum 2 choose treatment \(A\), resulting in cohort data collected from Stream 1. Before initiating the ``chosen'' treatment for each individual, we simulate Stream 2 as a random sample from the target population through a range of sampling rates, i.e., \(p_{2} = (5\%,\ 10\%,\ 20\%)\) and evenly assign the treatment at random to the individuals in it. All individuals who were not part of the random sample in Stream 2, or who were part of that sample and randomized to the same treatment that they chose in Stream 1, keep their initial assigned treatment. However, for the small contingent of ``unlucky'' ones, a switch is made to the other treatment (as per the
\textit{``label-switching''} strategy).

For the first simulation study assessing treatment effects with a binary outcome variable, we generated the treatment outcome (\(Y\)) such that 50\% of people in stratum 1 and 80\% of people in stratum 2 show a response (\(Y = 1\)) when using treatment \(A\). Conversely, 30\% of people in stratum 1 and 70\% of people in stratum 2 who receive treatment \(B\) demonstrate a response (\(Y = 1\)). With these specifications, the true outcome means (i.e., response proportions) from treatment \(A\) and \(B\) are 0.68 and 0.54 respectively, and the effect difference is 0.14. In the tables to follow, we evaluate results for a population size \(N_{tot} = 1,000\) based on 2,000 simulation runs per scenario. The proposed Bayesian credible interval is evaluated via 1,000 posterior samples in each iteration. A more expanded set of simulation scenarios examining population sizes of \(N_{tot} = 500\) and 5,000 can be found in the Appendix 4 of {\textit{Supplementary Materials}} (Tables S3-S8).

We compare the performance of each estimator in Table \ref{tbl.3}, with a focus on the treatment \(A\) response rate. As anticipated, estimation based on Stream 1 only (\({\hat{\mu}}_{1}\)) yields biased estimates due to the non-representative sampling; the mechanisms behind this would typically be unknown in practice. The other estimators yield negligible empirical bias as expected, benefitting from the anchor stream design. In each setting, both the CRC estimators \({\hat{\mu}}_{\hat{\Psi}}\) and \({\hat{\mu}}_{CRC}\) yield greater precision than the random sampling-based estimator (\({\hat{\mu}}_{RS}\)), and the Chapman estimator (\({\hat{\mu}}_{Chap}\)). In particular, the CRC estimator \({\hat{\mu}}_{\hat{\Psi}}\) demonstrates performance akin to that of the CRC estimator \({\hat{\mu}}_{CRC}\); however, the latter exhibits a slightly smaller standard error and narrower interval width, attributed to its utilization of the full set of observations in Table \ref{tbl.2}.

The Wald-type CIs of the CRC estimators \({\hat{\mu}}_{\hat{\Psi}}\) and \({\hat{\mu}}_{CRC}\) tend to be anti-conservative when the sampling rate (\(p_{2}\)) into Stream 2 is small. In contrast, the proposed Bayesian credible interval approach demonstrates a significant improvement in terms of the coverage of each interval, especially when \(p_{2} = 5\%\). This approach effectively accounts for the uncertainty in the CRC estimators and provides a stable credible interval across a wide range of the sampling rate.

\begin{table}[t]
\centering
\caption{Simulation result to compare the estimation for treatment $A$ with $\mu_{\text{true}} = 0.68, \\ \, N_{\text{tot}}=1,000$}
\label{tbl.3}
\renewcommand{\arraystretch}{1.3}
\setlength{\tabcolsep}{9pt}
\begin{tabular}{ccccccc}
\toprule
Setting & Estimation & $\hat{\mu}_{1}~^{a}$ & $\hat{\mu}_{RS}$ & $\hat{\mu}_{Chap}~^{b}$ & $\hat{\mu}_{\hat{\Psi}}~^{c}$ & $\hat{\mu}_{CRC}~^{c}$ \\
\midrule
\multirow{5}{*}{$p_{2}=5\%$} 
 & mean   & 0.752 & 0.678 & 0.673 & 0.678 & 0.677 \\
 & SD     & 0.019 & 0.097 & 0.149 & 0.072 & 0.074 \\
 & Avg.SE & 0.019 & 0.091 & 0.132 & 0.067 & 0.066 \\
 & Width  & 0.075 & 0.357 & 0.446 & 0.262 (\textbf{0.249}) & 0.258 (\textbf{0.234}) \\
 & CI (\%) & 4.9 & 89.7 & 97.3 & 90.6 (\textbf{95.4}) & 87.8 (\textbf{94.5}) \\
\midrule
\multirow{5}{*}{$p_{2}=10\%$} 
 & mean   & 0.751 & 0.681 & 0.679 & 0.681 & 0.680 \\
 & SD     & 0.019 & 0.064 & 0.098 & 0.049 & 0.050 \\
 & Avg.SE & 0.020 & 0.065 & 0.096 & 0.049 & 0.048 \\
 & Width  & 0.077 & 0.256 & 0.369 & 0.190 (\textbf{0.190}) & 0.189 (\textbf{0.179}) \\
 & CI (\%) & 4.7 & 92.8 & 95.8 & 94.1 (\textbf{95.6}) & 92.5 (\textbf{94.7}) \\
\midrule
\multirow{5}{*}{$p_{2}=20\%$} 
 & mean   & 0.751 & 0.681 & 0.679 & 0.680 & 0.680 \\
 & SD     & 0.020 & 0.047 & 0.070 & 0.037 & 0.037 \\
 & Avg.SE & 0.020 & 0.046 & 0.066 & 0.035 & 0.035 \\
 & Width  & 0.079 & 0.182 & 0.253 & 0.139 (\textbf{0.147}) & 0.139 (\textbf{0.134}) \\
 & CI (\%) & 6.9 & 94.2 & 85.1 & 94.1 (\textbf{94.3}) & 94.3 (\textbf{94.4}) \\
\bottomrule
\end{tabular}

\vspace{0.6em}
\raggedright
\footnotesize \textit{a. the estimation result based on Stream 1 only is reported for $\hat{\mu}_{1}$ \\
b. the transformed logit CI \citep{Sadinle2009} is reported for $\hat{\mu}_{Chap}$ \\
c. the proposed Bayesian Credible Interval (\textbf{bold}) is reported for $\hat{\mu}_{\hat{\Psi}}$ and $\hat{\mu}_{CRC}$}
\end{table}

\begin{table}[t]
\centering
\caption{Simulation result to compare the estimation for treatment $B$ with $\mu_{\text{true}} = 0.54, \\ \, N_{\text{tot}}=1,000$}
\label{tbl.4}
\renewcommand{\arraystretch}{1.3}
\setlength{\tabcolsep}{9pt}
\begin{tabular}{ccccccc}
\toprule
Setting & Estimation & $\hat{\mu}_{1}~^{a}$ & $\hat{\mu}_{RS}$ & $\hat{\mu}_{Chap}~^{b}$ & $\hat{\mu}_{\hat{\Psi}}~^{c}$ & $\hat{\mu}_{CRC}~^{c}$ \\
\midrule
\multirow{5}{*}{$p_{2}=5\%$} 
 & mean   & 0.443 & 0.538 & 0.524 & 0.539 & 0.540 \\
 & SD     & 0.029 & 0.101 & 0.252 & 0.084 & 0.084 \\
 & Avg.SE & 0.029 & 0.098 & 0.217 & 0.081 & 0.079 \\
 & Width  & 0.113 & 0.383 & 0.665 & 0.316 (\textbf{0.298}) & 0.308 (\textbf{0.285}) \\
 & CI (\%) & 7.9 & 92.2 & 97.1 & 92.1 (\textbf{95.7}) & 91.0 (\textbf{94.9}) \\
\hline
\multirow{5}{*}{$p_{2}=10\%$} 
 & mean   & 0.442 & 0.541 & 0.545 & 0.542 & 0.542 \\
 & SD     & 0.030 & 0.071 & 0.207 & 0.060 & 0.060 \\
 & Avg.SE & 0.029 & 0.070 & 0.171 & 0.058 & 0.057 \\
 & Width  & 0.115 & 0.273 & 0.572 & 0.228 (\textbf{0.223}) & 0.223 (\textbf{0.213}) \\
 & CI (\%) & 8.2 & 93.9 & 95.1 & 93.6 (\textbf{95.2}) & 92.6 (\textbf{94.6}) \\
\hline
\multirow{5}{*}{$p_{2}=20\%$} 
 & mean   & 0.442 & 0.541 & 0.539 & 0.540 & 0.540 \\
 & SD     & 0.030 & 0.049 & 0.129 & 0.043 & 0.043 \\
 & Avg.SE & 0.030 & 0.050 & 0.118 & 0.042 & 0.041 \\
 & Width  & 0.118 & 0.194 & 0.442 & 0.164 (\textbf{0.166}) & 0.162 (\textbf{0.157}) \\
 & CI (\%) & 9.7 & 94.7 & 91.1 & 93.9 (\textbf{94.7}) & 93.8 (\textbf{94.3}) \\
\bottomrule
\end{tabular}

\vspace{0.6em}
\raggedright
\footnotesize \textit{a. the estimation result based on Stream 1 only is reported for $\hat{\mu}_{1}$ \\
b. the transformed logit CI \citep{Sadinle2009} is reported for $\hat{\mu}_{Chap}$ \\
c. the proposed Bayesian Credible Interval (\textbf{bold}) is reported for $\hat{\mu}_{\hat{\Psi}}$ and $\hat{\mu}_{CRC}$}
\end{table}

Results pertaining to the estimated outcome mean for treatment \(B\) are presented in Table \ref{tbl.4}, leading to qualitatively similar conclusions to those based on Table \ref{tbl.3}. Given the closely aligned performance of the two CRC estimators \({\hat{\mu}}_{\hat{\Psi}}\) and \({\hat{\mu}}_{CRC}\), either could be recommended in practice. However, it is worth noting that the estimator \({\hat{\mu}}_{CRC}\) tends to provide a narrower Bayesian credible interval compared to \({\hat{\mu}}_{\hat{\Psi}}\).

\begin{table}[t]
\centering
\caption{Simulation result to compare the average treatment effect (ATE)$^a$ with $ATE_{\text{true}} = 0.14, \\ \, N_{\text{tot}}=1,000$}
\label{tbl.5}
\renewcommand{\arraystretch}{1.3}
\setlength{\tabcolsep}{9pt}
\begin{tabular}{ccccccc}
\toprule
Setting & Estimation & $ATE_{1}~^{b}$ & $ATE_{RS}$ & $ATE_{Chap}$ & $ATE_{\hat{\Psi}}~^{c}$ & $ATE_{CRC}~^{c}$ \\
\midrule
\multirow{5}{*}{$p_{2}=5\%$} 
 & Mean   & 0.309 & 0.139 & 0.150 & 0.139 & 0.138 \\
 & SD     & 0.035 & 0.140 & 0.292 & 0.112 & 0.113 \\
 & Avg.SE & 0.035 & 0.134 & 0.267 & 0.105 & 0.103 \\
 & Width  & 0.136 & 0.524 & 1.047 & 0.412 (\textbf{0.393}) & 0.404 (\textbf{0.372}) \\
 & CI (\%) & 0.2 & 93.1 & 97.5 & 93.0 (\textbf{95.0}) & 91.5 (\textbf{94.2}) \\
\hline
\multirow{5}{*}{$p_{2}=10\%$} 
 & Mean   & 0.309 & 0.140 & 0.134 & 0.139 & 0.139 \\
 & SD     & 0.036 & 0.094 & 0.231 & 0.076 & 0.077 \\
 & Avg.SE & 0.035 & 0.096 & 0.203 & 0.076 & 0.075 \\
 & Width  & 0.138 & 0.374 & 0.794 & 0.297 (\textbf{0.294}) & 0.293 (\textbf{0.278}) \\
 & CI (\%) & 0.3 & 95.3 & 95.2 & 94.9 (\textbf{95.5}) & 93.5 (\textbf{94.7}) \\
\hline
\multirow{5}{*}{$p_{2}=20\%$} 
 & Mean   & 0.310 & 0.140 & 0.140 & 0.140 & 0.140 \\
 & SD     & 0.037 & 0.068 & 0.147 & 0.055 & 0.055 \\
 & Avg.SE & 0.036 & 0.068 & 0.137 & 0.055 & 0.054 \\
 & Width  & 0.141 & 0.266 & 0.538 & 0.215 (\textbf{0.223}) & 0.213 (\textbf{0.207}) \\
 & CI (\%) & 0.5 & 95.1 & 94.4 & 94.6 (\textbf{94.7}) & 94.2 (\textbf{94.5}) \\
\bottomrule
\end{tabular}

\vspace{0.6em}
\raggedright
\footnotesize \textit{a. the average treatment effects (ATE) equals to $\hat{\mu}_A - \hat{\mu}_B$ \\
b. the estimation result based on Stream 1 only is reported for $\hat{\mu}_{1}$ \\
c. the proposed Bayesian Credible Interval (\textbf{bold}) is reported for ${ATE}_{\hat{\Psi}}$ and ${ATE}_{CRC}$}
\end{table}

In Table \ref{tbl.5}, we evaluate the Average Treatment Effect (ATE) based on the outcome means given in Table \ref{tbl.3} and Table \ref{tbl.4}. As expected, the estimate from the Stream 1 data is still biased due to its non-representativeness, while the other estimators are essentially unbiased. Notably, \({ATE}_{Chap}\) suffers from slight bias here, as it loses some estimation accuracy due to zero-counts in some cells of Table \ref{tbl.1} when the sampling rate of stream 2 is small. However, the more serious problem is that Chapman's estimator is highly inefficient in this setting. Overall, the proposed CRC estimators, \({ATE}_{\hat{\Psi}}\) and \({ATE}_{CRC}\), demonstrate the best performance for ATE estimation. In particular, \({ATE}_{CRC}\), together with the proposed Bayesian credible interval, provides the most reliable and precise estimation.

For the second simulation study to investigate treatment effects in terms of general means, we generated a continuous outcome \(\widetilde{Y}\) characterized by heterogeneity in its distribution across members of the simulated target population. We adopt a mixture of varying normal distributions based on different strata, treatment binary response (\(Y\)) and treatment selection. Specifically, the continuous outcome \(\widetilde{Y}\) is generated from eight different normal distributions with the combination of \((i,\ j,\ k,\ \mu,\ \sigma)\), where \(i = 1,\ 2\) for strata, \(j = 1,\ 0\) for treatment binary response \(Y\), \(k = A,\ B\) for treatment selection, and mean and standard deviation \(\mu\), \(\sigma\): (1, 1, \(A\), 10, 0.75), (1, 0, \(A\), 2.5, 1.2), (2, 1, \(A\), 5, 0.5), (2, 0, \(A\), 1, 1.5), (1, 1, \(B\), 15, 0.75), (1, 0, \(B\), 7.5, 1.2), (2, 1, \(B\), 10, 0.5), (2, 0, \(B\), 6, 1.5). Based on the weighted average of each normal distribution, the true overall mean of \(\widetilde{Y}\) is therefore calculated as \(\mu_{A} = 5.02\), \(\mu_{B} = 9.18\). The true mean difference (treatment effect) follows, i.e., \(\mu^{AB} = \mu_{A} - \mu_{B} = - 4.16\).

The results of this simulation study with a population size (\(N_{tot} = 1,000\)) and sampling rate for Stream 2 (\(p_{2} = 10\%\)) are summarized in Table \ref{tbl.6}. We examined the proposed mean estimators based on (\ref{eq.8}) and the treatment difference (ATE) between treatment groups \(A\) and \(B\). For each estimator, we compared three distinct methods for assessing the mean of the continuous outcome \(\widetilde{Y}\). The ``Stream 1 only'' method derives the mean estimate solely from Stream 1 data. All estimates calculated in this way are biased due to the nonrepresentative sampling scheme of Stream 1. In contrast, the ``Stream 2 only'' method calculates the mean estimate exclusively from the anchor stream (Stream 2), yielding unbiased results as anticipated. Meanwhile, incorporating both Stream 1 and Stream 2 data, the ``CRC'' method reports more efficient mean estimates based on the capture-recapture framework. A more expanded set of simulation scenarios examining different sampling rate of Stream 2 (\(p_{2} = 5\%,\ 20\%\)) can be found in the Appendix 4 of {\textit{Supplementary Materials}} (Tables S9-S10).

\begin{table}[ht]
\centering
\caption{Simulations Evaluating Mean Estimates for Continuous $X$ with $N_{\text{tot}}=1000, \, p_{2}=10\%$}
\label{tbl.6}
\renewcommand{\arraystretch}{1.3}
\setlength{\tabcolsep}{8pt}
\begin{tabular}{cccccccc}
\toprule
\multirow{2}{*}{Estimator} & {True} & \multirow{2}{*}{Methods} & \multirow{2}{*}{Mean} & \multirow{2}{*}{SD} & {Average} & {CI} & {Average} \\
 & mean & & & & SE $^a$ & Coverage\% & CI Width \\
\hline
\multirow{3}{*}{$\hat{\mu}_A~^{b}$} 
 & \multirow{3}{*}{5.020} 
 & Stream 1 only & 4.533 & 0.110 & -- & -- & -- \\
 &  & Stream 2 only & 5.036 & 0.431 & 0.423 & 94.4 & 1.646 \\
 &  & CRC & 5.033 & 0.359 & 0.347 & 93.5 & 1.344 \\
\hline
\multirow{3}{*}{$\hat{\mu}_B~^{b}$} 
 & \multirow{3}{*}{9.180} 
 & Stream 1 only & 9.414 & 0.190 & -- & -- & -- \\
 &  & Stream 2 only & 9.178 & 0.387 & 0.397 & 93.9 & 1.544 \\
 &  & CRC & 9.181 & 0.323 & 0.316 & 93.1 & 1.231 \\
\hline
\multirow{3}{*}{$\hat{\mu}^{AB}$} 
 & \multirow{3}{*}{-4.160} 
 & Stream 1 only & -4.880 & 0.216 & -- & -- & -- \\
 &  & Stream 2 only & -4.141 & 0.586 & 0.581 & 94.6 & 2.262 \\
 &  & CRC & -4.147 & 0.493 & 0.471 & 93.6 & 1.832 \\
\bottomrule
\end{tabular}

\vspace{0.6em}
\raggedright
\footnotesize \textit{a. SE for each estimator based on bootstrap with percentile CIs. \\
b. SE, CIs and their widths for the estimated mean not reported for the estimator based on Stream 1 only.}
\end{table}

\section{Illustrative Data Example}\label{sec4}

The design and estimation approaches outlined above demonstrated clear advantages for strengthening treatment effect evaluation in observational cohort studies, both conceptually and empirically, as shown through the simulation studies. However, implementing such approaches in practice requires careful adherence to strict guidelines for random sampling and \emph{``label-switching''}. Given our proposed CRC framework for treatment effect evaluation, we present an illustrative data example using two research studies comparing the antibody response to two Covid-19 vaccines.

Beginning in early 2020, a newly discovered coronavirus, Severe Acute Respiratory Syndrome Coronavirus 2 (SARS-CoV-2), spread worldwide. In response, healthcare experts and pharmaceutical companies worked collaboratively to develop vaccines to combat the virus. To date, numerous studies \citep{BenAhmed2022,Jeewandara2022,Mok2022} have compared different vaccines, focusing on major Covid-19 vaccines such as mRNA-1273 (Moderna), BNT162B2 (Pfizer-BioNTech), Sputnik V (Gamaleya Research Institute), ChAdOx1-S (AstraZeneca), Sinopharm (BIBP), and Sinovac (Beijing). Most of these studies \citep{Jeewandara2022,Mok2022} are observational, examining vaccine effectiveness, efficacy, and antibody responses (seropositivity). However, the generalizability of these findings is often questioned due to selection bias. A smaller number of studies include randomized trials, although their limited sample sizes present challenges.

This article introduces a method for integrating data from both study types to strengthen treatment effect evaluation. We illustrate this approach by mimicking a randomized trial from Tunisia \citep{BenAhmed2022} and generating synthetic observational data purportedly from the same target population under a capture-recapture framework. In this example, we compare antibody responses following two doses of hypothetical treatment modeled after the Sputnik V and Sinopharm vaccines.

As of 12 January 2022, over 6 million individuals in Tunisia had completed vaccination with one of the common Covid-19 vaccines \citep{BenAhmed2022}. The synthetic target population for this example comprises 2,000 hypothetical Tunisians aged 40 and older who had not experienced symptomatic Covid-19 and had provided informed consent for vaccination. Following the study design outlined above, we assume participants received the vaccine of their choice or as recommended by their medical providers, after which their humoral antibody responses (Anti-S Antibodies) were assessed using a commercial Anti-SARS-CoV-2 test following the second dose. For this demonstration, test results on a random sample from the actual Tunisian target population (including 169 participants \citep{BenAhmed2022}, representing 8.45\% of the 2,000 individuals) were used to set parameters for generating anchor stream data to be combined with the synthetic observational data. We use synthetic individual-level data to emulate the target population in Tunisia by randomly sampling data until acquiring 85 seropositive responses for \(A\) and 71 for \(B\) among 169 random samples to mimic Stream 2 data based on \cite{BenAhmed2022}, serving as the anchor stream. Table \ref{tbl.7} presents the number and percentage of seropositive individuals in these empirical studies.

\begin{table}[ht]
\centering
\caption{Number and percent of seropositive participants for two vaccine types in Tunisia}\label{tbl.7}
\renewcommand{\arraystretch}{1.3}
\setlength{\tabcolsep}{9pt}
\begin{tabular}{cccccc}
\toprule
\multirow{2}{*}{Vaccine} & \multicolumn{2}{c}{Stream 1 $^{a}$} & & \multicolumn{2}{c}{Stream 2 $^{b}$} \\
\cline{2-3}
\cline{5-6}
 & $N$ & Seropositive (\%) &  & $N$ & Seropositive (\%) \\
\hline
Vaccine A & 327 & 293 (89.6\%) & &  86 & 85 (98.8\%) \\
Vaccine B & 571 & 508 (89.0\%) & & 83 & 71 (85.5\%) \\
\hline
Total     & 898 & -- & & 169 & -- \\
\bottomrule
\end{tabular}

\vspace{0.6em}
\raggedright
\footnotesize \textit{a. Stream 1 comprises synthetic data from observational cohorts. The total target population size is $N=2,000$. \\
b. Stream 2 mimics randomized trial cohorts from \cite{BenAhmed2022}. Vaccine $A$ represents Sputnik V, while Vaccine $B$ represents Sinopharm.}
\end{table}

For a synthetic target population, we assume that vaccine selection was associated with a variable such as insurance type, where older individuals and/or those of lower socioeconomic status might be covered by one type of insurance (comprising 5\% of the population, with 90\% receiving Vaccine \(A\) showing a seropositivity rate of 75\%, and the rest receiving Vaccine \(B\) with a seropositivity rate of 60\%). In contrast, younger or more affluent individuals might use another type of insurance (comprising 95\% of the population, with 30\% receiving Vaccine \(A\) showing a seropositivity rate of 99\%, and the remainder receiving Vaccine \(B\) with a seropositivity rate of 90\%). As a result, the true seropositivity rates were set at 97.8\% for Vaccine \(A\) and 88.5\% Vaccine \(B\). In the synthetic observational cohort (Stream 1), the vast majority (99\%) of subjects with the first type of insurance participated, while those with the second type had a lower participation rate (45\%), reflecting a logical source of potential non-representativeness.

\begin{table}[ht]
\centering
\caption{Seropositivity Estimates and Comparison for the Synthetic Population in Tunisia $^{a}$}\label{tbl.8}
\renewcommand{\arraystretch}{1.3}
\setlength{\tabcolsep}{7pt}
\begin{tabular}{cccccc}
\hline
Vaccine & Estimator & Mean & SE & 95\% CI $^{b}$ & Width \\
\hline
\multirow{2}{*}{Vaccine A} 
 & $\hat{\mu}_{RS,A}$  & 98.8\% & 0.0116 & [96.6\%, 100.0\%] & 0.034 $^{c}$ \\
 & $\hat{\mu}_{CRC,A}$ & 98.0\% & 0.0059 & [96.8\%, 99.1\%], \textbf{[95.4\%, 98.4\%]} & 0.023, \textbf{0.029} \\
\hline
\multirow{2}{*}{Vaccine B} 
 & $\hat{\mu}_{RS,B}$  & 85.5\% & 0.0386 & [78.0\%, 93.1\%] & 0.151 \\
 & $\hat{\mu}_{CRC,B}$ & 88.5\% & 0.0292 & [82.8\%, 94.3\%], \textbf{[81.8\%, 93.0\%]} & 0.114, \textbf{0.112} \\
\hline
\multirow{3}{*}{Difference $^{d}$} 
 & $ATE_{1}$    & 0.6\%  & 0.0213 & [0.0\%, 4.8\%] & 0.048 \\
 & $ATE_{RS}$   & 13.3\% & 0.0403 & [5.4\%, 21.2\%] & 0.158 \\
 & $ATE_{CRC}$  & 9.5\%  & 0.0298 & [3.6\%, 15.3\%], \textbf{[3.7\%, 16.4\%]} & 0.117, \textbf{0.127} \\
\hline
\end{tabular}

\vspace{0.6em}
\raggedright
\footnotesize \textit{a. In this example, the true seropositivity rates were set at 97.8\% for Vaccine A and 88.5\% Vaccine B. \\
b. The Wald-based CIs reported for $\hat{\mu}_{RS,A}$ and $\hat{\mu}_{RS,B}$ are based on the random sampling estimator given in (\ref{eq.1}); The Wald-based CIs reported for $\hat{\mu}_{CRC,A}$ and $\hat{\mu}_{CRC,B}$ are based on the variance estimator for (\ref{eq.3}) and (\ref{eq.4}) given in Appendix 3; The proposed Bayesian Credible Intervals (\textbf{bold}) reported for $\hat{\mu}_{CRC,A}$ and $\hat{\mu}_{CRC,B}$ are based on Section \ref{sec2.4}. \\
c. The upper limit of the CI for $\hat{\mu}_{RS,A}$ is capped at 100\%, with a width of 0.045 when disregarding the cap. \\
d. The difference of two vaccines equals to $\hat{\mu}_{RS,A} - \hat{\mu}_{RS,B}$ or $\hat{\mu}_{CRC,A} - \hat{\mu}_{CRC,B}$, which matches the definition of $ATE$ introduced in previous sections. $ATE_{1}$ is reported for the comparison based on Stream 1 only.}
\end{table}

For illustration, a single set of observed cell counts (see Table \ref{tbl.2}) was simulated as follows: \(n_{1}\)=12, \(n_{2}\)=1,
\(n_{3}\)=281, \(n_{4}\)=33, \(n_{5}\)=33, \(n_{6}\)=0, \(n_{7}\)=18,
\(n_{8}\)=5, \(n_{9}\)=490, \(n_{10}\)=58, \(n_{11}\)=14, \(n_{12}\)=2,
\(n_{13}\)=40, \(n_{14}\)=0, \(n_{15}\)=39, \(n_{16}\)=5,
\(n_{17}\)=969, corresponding to the counts in Table \ref{tbl.7}, e.g., the cell count for positive responses to Vaccine \(A\) in Stream 1 is
\(n_{S1}^{A} = n_{1} + n_{3} = 293\), and in Stream 2 it is
\(n_{S2}^{A} = n_{1} + n_{5} + n_{13} = 85\). We then compare the CRC estimator \({\hat{\mu}}_{CRC}\) based on (\ref{eq.3}) and (\ref{eq.4}) to the
random sampling estimator \({\hat{\mu}}_{RS}\) derived from the randomized trial data. The results of this example are presented in Table \ref{tbl.8}. Vaccine \(A\) and Vaccine \(B\), representing the two Covid-19 vaccines mentioned earlier, illustrate seropositivity estimates within the synthetic population. As anticipated, leveraging additional information from the observational cohorts provides significant benefits, such as reducing interval widths by approximately 20\% (e.g., from 0.158 to 0.127 for \({ATE}_{CRC}\)), consistent with conclusions drawn from the simulation studies. Furthermore, the difference in seropositivity rates estimated based on Stream 1 (\({ATE}_{1})\) is biased toward the null, clearly highlighting the common issue of selection bias inherent in observational data. Although only a single set of simulated data is presented, the Stream 2 data are exactly
representative of the Tunisian trial \citep{BenAhmed2022}. A reliability analysis is also provided in Appendix 5 of {\textit{Supplementary Materials}}, to demonstrate the robustness of this example as it applies to real-world data practice.

Additionally, most antibody comparison studies \citep{BenAhmed2022,Jeewandara2022,Mok2022} focus on comparing both the binary seropositivity response and continuous measurements of antibody levels, such as cellular immune
responses based on CD4 or CD8 levels. The extension introduced in Section \ref{sec2.5} is well suited for this context. However, due to the lack of relevant data, our illustrative example only addresses comparisons of seropositive response rate (binary outcomes). Given availability of continuous outcome data, the proposed method can be readily applied to address such research interests.

\section{Discussion}\label{sec5}

In this article, we have employed capture-recapture methods to evaluate treatment effects and enhance inference about a trial-eligible target population within an observational cohort. We have introduced several estimators to evaluate the response probabilities for the individual treatments, as well as the average treatment effect (ATE). Our empirical studies suggest that the proposed anchor stream-based estimators provide unbiased and efficient estimation for the outcome mean of a single treatment as well as the ATE, with enhanced precision compared to the random sampling-based estimator. As an application, we demonstrated our method using an illustrative example based on a randomized trial from Tunisia, comparing Anti-S Antibody seropositive response rates between two major Covid-19 vaccines and yielding conclusions consistent with those from the empirical studies. All R programs related to the simulation studies and the illustrative data example are available on GitHub (\url{https://github.com/lge-biostat/CRC_treatment_effects}).

The proposed approach to performing the CRC analysis builds on the study design introduced in Section 1 and requires the capacity to draw a representative sample from a well-defined target population that consists of a list or registry of individuals eligible for treatment assessment. Similar study designs exist for combining randomized trials and observational studies, such as pragmatic randomized trials nested within a cohort of eligible individuals \citep{Ford2016,newman2016}. These designs assume that the observational data can provide a good representative basis for the randomized trial, and that treatment effect evaluation based on an embedded randomized trial could improve the generalizability and transportability. In contrast, our approach does not rely on crucial assumptions about the nature of the non-representativeness of observational study participants, as these are often unverifiable in practice \citep{Hammer2009}. Instead, it requires only the basic treatment consistency assumption as a minimal condition. Building on this foundation, we leverage what can be a much smaller representative random sample from the target population to anchor the estimation validity, while borrowing added precision from the observational component. The proposed anchor stream-based CRC estimators leverage the generalizability of the representative sample and ``transport'' the observational information, thereby enhancing estimation precision. This approach may offer a novel strategy within the field of causal inference when the design is feasible.

Future work may consider extending the treatment evaluation from the trial-eligible target population to a more general target population. To achieve this goal, baseline covariate information may need to be considered, and a stratified sampling approach to acquire the anchor stream may be necessary. We also anticipate generalization of the measure of treatment effect considered here, as well as potential efforts to target favorable bias-variance tradeoffs if covariates deemed to explain the majority of the observational non-representativeness are available.

% \section{Supplementary Material}
% \label{sec6}

% Supplementary material is available online at
% % \href{http://biostatistics.oxfordjournals.org}%
% % {http://biostatistics.oxfordjournals.org}.
% \url{http://biostatistics.oxfordjournals.org}.

\section*{Acknowledgments}

This work was supported by the National Institute of Health (NIH)/National Institute of Allergy and Infectious Diseases (P30AI050409; Del Rio PI), the NIH/National Center for Advancing Translational Sciences (UL1TR002378; Taylor PI), the NIH/National Cancer Institute (R01CA234538; Ward/Lash MPIs), and the NIH/National Cancer Institute (R01CA266574; Lyles/Waller MPIs).

{\it Conflict of Interest}: None declared.

\bibliographystyle{biorefs}
\bibliography{references}

\end{document}

% Title of paper
\title{A Capture-Recapture Approach to Enhance Treatment Effect Evaluation in an Observational Cohort - Supplementary Materials}

% List of authors, with corresponding author marked by asterisk
\author{Lin Ge$^1$, Yuzi Zhang$^2$, Lance A. Waller$^3$, Robert H. Lyles$^3$\\[4pt]
% Author addresses
\textit{$^1$Department of Epidemiology and Biostatistics, School of Public Health,
Indiana University, Bloomington, IN, U.S.A
$^2$Division of Biostatistics, College of Public Health, Ohio State University, Columbus, OH, U.S.A
$^3$Department of Biostatistics and Bioinformatics, Rollins School of Public Health,
Emory University, Atlanta, GA, U.S.A
}
\\[2pt]
% E-mail address for correspondence
{lge\_biostat@outlook.com}}

% Running headers of paper:
\markboth%
% First field is the short list of authors
{L. Ge and others}
% Second field is the short title of the paper
{Capture-Recapture Approach to Enhance Treatment Effect Evaluation}

\maketitle

\section*{Appendix 1 Derivation of Likelihood Contributions for Table 2}

For each observation type in Table 2, we first derive the likelihood contribution for each cell. As examples, consider cells \(n_{1}\), \(n_{3}\), \(n_{5}\), and \(n_{13}\) corresponding to receiving
treatment \(A\).

\begin{itemize}
\item
  For \(n_{1}\): the observation type \\
  \textit{\{sampled in both streams, assigned and randomized to $A$, $Y=1$\}}\\
  is equivalent to\\
  \emph{\{sampled in S1} \(\cap\) \emph{assigned} \(A\) \emph{in
S1\}}\(\cap\)\emph{\{sampled in S2} \(\cap \ \)\emph{randomized to}
\(A\) \emph{in S2\}}\(\cap\)
\emph{\{received} \(A\)\emph{\}}\(\cap\)\emph{\{}$Y=1$\emph{\} =} \((E_{S1} \cap E_{S1,A}) \cap (E_{S2} \cap E_{S2,A}) \cap E_{A} \cap \{ Y = 1 \}\),\\
where \(E_{S1} = \{ sampled\ in\ S1\}\),
\(E_{S1,A} = \left\{ assigned\ A\ in\ S1 \right\}\),
\(E_{S2} = \{ sampled\ in\ S2\}\),
\(E_{S1,A} = \{ randomized\ to\ A\ in\ S2\}\)\emph{.}

In addition, the event \(E_{A} =\)\emph{\{received} \(A\)\emph{\}} is equivalent to \\
\emph{\{not sampled in S2} \(\cap\) \emph{sampled in S1}
\(\cap \ \)\emph{assigned} \(A\) \emph{in S1\}}\(\cup\) \emph{\{sampled in S2} \(\cap \ \)\emph{randomized to} \(A\)
\emph{in S2\} =}
\(({\overline{E}}_{S2} \cap E_{S1} \cap E_{S1,A}) \cup (E_{S2} \cap E_{S2,A})\),\\
where \({\overline{E}}_{S2} = \{ not\ sampled\ in\ S2\}\)\emph{.}

Therefore, the likelihood contribution for \(n_{1}\) can be written as follows.

\(p_{1}\) = Pr{[}\emph{\{sampled in S1} \(\cap \ \)\emph{assigned} \(A\)
\emph{in S1\}}\(\cap\)\emph{\{sampled in S2} \(\cap \ \)\emph{randomized
to} \(A\) \emph{in S2\}} \(\cap\) \emph{\{received}
\(A\)\emph{\}}\(\cap\)\emph{\{}$Y=1$\emph{\}}{]}

=
Pr{[}\((E_{S1} \cap E_{S1,A}) \cap (E_{S2} \cap E_{S2,A}) \cap E_{A} \cap \{ Y=1 \}\){]}

= Pr{[}$Y=1$ \emph{\textbar{}}
\((E_{S1} \cap E_{S1,A}) \cap (E_{S2} \cap E_{S2,A}) \cap E_{A}\){]}
\(\times\) Pr{[}\((E_{S1} \cap E_{S1,A}) \cap E_{A}\) \emph{\textbar{}}
\(E_{S2} \cap E_{S2,A}\){]} \(\times \ \)Pr{[}\(E_{S2,A}\)
\emph{\textbar{}} \(E_{S2}\){]} \(\times\) Pr{[}\(E_{S2}\){]}

= Pr{[}$Y=1$
\emph{\textbar{}}\((E_{S1} \cap E_{S1,A}) \cap E_{A}\){]} \(\times\)
Pr{[}\(E_{S1} \cap E_{S1,A}\) \emph{\textbar{}}
\(E_{S2} \cap E_{S2,A}\){]} \(\times\) Pr{[}\(E_{A}\) \emph{\textbar{}}
\(E_{S2} \cap E_{S2,A}\){]} \(\times \ \)Pr{[}\(E_{S2,A}\)
\emph{\textbar{}} \(E_{S2}\){]} \(\times\) Pr{[}\(E_{S2}\){]}

= Pr{[}$Y=1$
\emph{\textbar{}}\((E_{S1} \cap E_{S1,A}) \cap E_{A}\){]} \(\times\)
Pr{[}\(E_{S1} \cap E_{S1,A}\){]} \(\times\) 1 \(\times\)
Pr{[}\(E_{S2,A}\) \emph{\textbar{}} \(E_{S2}\){]} \(\times\)
Pr{[}\(E_{S2}\){]}

= Pr{[}$Y=1$
\emph{\textbar{}}\((E_{S1} \cap E_{S1,A}) \cap E_{A}\){]} \(\times\)
Pr{[}\(E_{S1,A}\) \emph{\textbar{}} \(E_{S1}\){]} \(\times\)
Pr{[}\(E_{S1}\){]} \(\times \ \)Pr{[}\(E_{S2,A}\) \emph{\textbar{}}
\(E_{S2}\){]} \(\times\) Pr{[}\(E_{S2}\){]}

= \(\pi_{s1,A}\phi_{A}\phi\xi_{A}\psi\)\\
where we define the parameters \(\pi_{s1,A}\) = Pr{[}$Y=1$
\emph{\textbar{}}\((E_{S1} \cap E_{S1,A}) \cap E_{A}\){]} =
Pr($Y=1$ \textbar{} sampled and assigned \(A\) in S1, and
received \(A\)), \(\phi_{A}\) = Pr{[}\(E_{S1,A}\) \emph{\textbar{}}
\(E_{S1}\){]} = Pr(Assigned \(A\) in S1 \textbar{} sampled in S1),
\(\phi\) = Pr{[}\(E_{S1}\){]} = Pr(Sampled in S1), \(\xi_{A}\) =
Pr{[}\(E_{S2,A}\) \emph{\textbar{}} \(E_{S2}\){]} = Pr(Randomized to
\(A\) \textbar{} sampled in S2) and\(\ \psi\) = Pr{[}\(E_{S2}\){]}=
Pr(Sampled in S2).

\item
  For \(n_{3}\): the observation type\\
  \emph{\{sampled and assigned treatment} \(A\) \emph{in S1, but not sampled in S2,} $Y=1$\emph{\}}\\
  is equivalent to\\
  \emph{\{sampled in S1} \(\cap \ \)\emph{assigned} \(A\) \emph{in
S1\}}\(\cap\)\emph{\{not sampled in S2\}}\(\cap\)\emph{\{received}
\(A\)\emph{\}}\(\cap\)\{$Y=1$\} \\
\emph{=}
\((E_{S1} \cap E_{S1,A}) \cap {\overline{E}}_{S2} \cap E_{A} \cap \{Y=1\}\)\emph{.}

Then the likelihood contribution is written as follows.

\(p_{3}\) =
Pr{[}\((E_{S1} \cap E_{S1,A}) \cap {\overline{E}}_{S2} \cap E_{A} \cap \{Y=1\}\){]}

= Pr{[}$Y=1$ \emph{\textbar{}}
\((E_{S1} \cap E_{S1,A}) \cap {\overline{E}}_{S2} \cap E_{A}\){]}
\(\times\) Pr{[}\((E_{S1} \cap E_{S1,A}) \cap E_{A}\) \emph{\textbar{}}
\({\overline{E}}_{S2}\){]} \(\times \ \)Pr{[}\({\overline{E}}_{S2}\){]}

= Pr{[}$Y=1$
\emph{\textbar{}}\((E_{S1} \cap E_{S1,A}) \cap E_{A}\){]} \(\times\)
Pr{[}\(E_{S1} \cap E_{S1,A}\){]} \(\times \ \)(1 - Pr{[}\(E_{S2}\){]})

= Pr{[}$Y=1$
\emph{\textbar{}}\((E_{S1} \cap E_{S1,A}) \cap E_{A}\){]} \(\times\)
Pr{[}\(E_{S1,A}\) \emph{\textbar{}} \(E_{S1}\){]} \(\times\)
Pr{[}\(E_{S1}\){]} \(\times \ \)(1 - Pr{[}\(E_{S2}\){]})

= \(\pi_{s1,A}\phi_{A}\phi(1 - \psi)\)

\item
  For \(n_{5}\): the observation type\\
  \emph{\{sampled and assigned treatment B in S1; sampled in S2, randomized and switched label to} \(A\), $Y=1$\emph{\}}\\
  is equivalent to\\
  \emph{\{sampled in S1} \(\cap\) \emph{assigned} \(B\) \emph{in
S1\}} \(\cap\) \emph{\{sampled in S2} \(\cap \ \)\emph{randomized to}
\(A\) \emph{in S2\}} \(\cap\) \emph{\{received} \(A\)\emph{\}} \(\cap\) \{$Y=1$\} \emph{=}
\((E_{S1} \cap E_{S1,B}) \cap (E_{S2} \cap E_{S2,A}) \cap E_{A} \cap \{Y=1\}\)\emph{.}

Then the likelihood contribution is written as follows.\\
\(p_{5}\) =
Pr{[}\((E_{S1} \cap E_{S1,B}) \cap (E_{S2} \cap E_{S2,A}) \cap E_{A} \cap \{Y=1\}\){]}

= Pr{[}$Y=1$ \emph{\textbar{}}
\((E_{S1} \cap E_{S1,B}) \cap (E_{S2} \cap E_{S2,A}) \cap E_{A}\){]}
\(\times\) Pr{[}\((E_{S1} \cap E_{S1,B}) \cap E_{A}\) \emph{\textbar{}}
\(E_{S2} \cap E_{S2,A}\){]} \(\times \ \)Pr{[}\(E_{S2,A}\)
\emph{\textbar{}} \(E_{S2}\){]} \(\times\) Pr{[}\(E_{S2}\){]}

= Pr{[}$Y=1$
\emph{\textbar{}}\((E_{S1} \cap E_{S1,B}) \cap E_{A}\){]} \(\times\)
Pr{[}\(E_{S1} \cap E_{S1,B}\) \emph{\textbar{}}
\(E_{S2} \cap E_{S2,A}\){]} \(\times\) Pr{[}\(E_{A}\) \emph{\textbar{}}
\(E_{S2} \cap E_{S2,A}\){]} \(\times \ \)Pr{[}\(E_{S2,A}\)
\emph{\textbar{}} \(E_{S2}\){]} \(\times\) Pr{[}\(E_{S2}\){]}

= Pr{[}$Y=1$
\emph{\textbar{}}\((E_{S1} \cap E_{S1,B}) \cap E_{A}\){]} \(\times\)
Pr{[}\(E_{S1} \cap E_{S1,B}\){]} \(\times\) 1 \(\times\)
Pr{[}\(E_{S2,A}\) \emph{\textbar{}} \(E_{S2}\){]} \(\times\)
Pr{[}\(E_{S2}\){]}

= Pr{[}$Y=1$
\emph{\textbar{}}\((E_{S1} \cap E_{S1,B}) \cap E_{A}\){]} \(\times\)
Pr{[}\(E_{S1,B}\) \emph{\textbar{}} \(E_{S1}\){]} \(\times\)
Pr{[}\(E_{S1}\){]} \(\times \ \)Pr{[}\(E_{S2,A}\) \emph{\textbar{}}
\(E_{S2}\){]} \(\times\) Pr{[}\(E_{S2}\){]}

= \(\pi_{s1\_ B,A}(1 - \phi_{A})\phi\xi_{A}\psi\)

where we define a new parameter \(\pi_{s1\_ B,A}\) = Pr{[}$Y=1$
\emph{\textbar{}}\((E_{S1} \cap E_{S1,B}) \cap E_{A}\){]} =
Pr($Y=1$ \textbar{} sampled and assigned \(B\) in S1, but
received \(A\)).

\item
  For \(n_{13}\): the observation type\\
  \emph{\{not sampled in S1; sampled in S2, randomized and switched label to} \(A\), $Y=1$\emph{\}}\\
  is equivalent to\\
  \emph{\{not sampled in S1\}} \(\cap\) \emph{\{sampled in S2}
\(\cap \ \)\emph{randomized to} \(A\) \emph{in S2\}} \(\cap\) \emph{\{received} \(A\)\emph{\}} \(\cap\) \{$Y=1$\} 
\emph{=}\({\overline{E}}_{S1} \cap (E_{S2} \cap E_{S2,A}) \cap E_{A} \cap \{Y=1\}\)\emph{.}\\
Then the likelihood contribution is written as follows.\\
\(p_{13}\) =
Pr{[}\({\overline{E}}_{S1} \cap (E_{S2} \cap E_{S2,A}) \cap E_{A} \cap \{Y=1\}\){]}\\
= Pr{[}$Y=1$ \emph{\textbar{}}
\({\overline{E}}_{S1} \cap (E_{S2} \cap E_{S2,A}) \cap E_{A}\){]}
\(\times\) Pr{[}\({\overline{E}}_{S1} \cap E_{A}\) \emph{\textbar{}}
\(E_{S2} \cap E_{S2,A}\){]} \(\times \ \)Pr{[}\(E_{S2,A}\)
\emph{\textbar{}} \(E_{S2}\){]} \(\times\) Pr{[}\(E_{S2}\){]}

= Pr{[}$Y=1$
\emph{\textbar{}}\({\overline{E}}_{S1} \cap E_{A}\){]} \(\times\)
Pr{[}\({\overline{E}}_{S1}\)\emph{\textbar{}}
\(E_{S2} \cap E_{S2,A}\){]} \(\times\) Pr{[}\(E_{A}\) \emph{\textbar{}}
\(E_{S2} \cap E_{S2,A}\){]} \(\times \ \)Pr{[}\(E_{S2,A}\)
\emph{\textbar{}} \(E_{S2}\){]} \(\times\) Pr{[}\(E_{S2}\){]}

= Pr{[}$Y=1$
\emph{\textbar{}}\({\overline{E}}_{S1} \cap E_{A}\){]} \(\times\)
Pr{[}\({\overline{E}}_{S1}\){]} \(\times\) 1 \(\times\)
Pr{[}\(E_{S2,A}\) \emph{\textbar{}} \(E_{S2}\){]} \(\times\)
Pr{[}\(E_{S2}\){]}

= Pr{[}$Y=1$
\emph{\textbar{}}\({\overline{E}}_{S1} \cap E_{A}\){]} \(\times\) (1-
Pr{[}\(E_{S1}\){]}) \(\times \ \)Pr{[}\(E_{S2,A}\) \emph{\textbar{}}
\(E_{S2}\){]} \(\times\) Pr{[}\(E_{S2}\){]}

= \(\pi_{s1\_ NA,A}(1 - \phi)\xi_{A}\psi\)

where we define a new parameter \(\pi_{s1\_ NA,A}\) = Pr{[}$Y=1$
\emph{\textbar{}}\({\overline{E}}_{S1} \cap E_{A}\){]} =
Pr($Y=1$ \textbar{} not sampled in S1, but received \(A\)).
  
\end{itemize}

The derivation for the remaining cells follows a similar procedure. In addition to the parameters for treatment \(A\), three new parameters are defined for treatment \(B\): \(\pi_{s1,B}\), \(\pi_{s1\_ A,B}\), and \(\pi_{s1\_ NA,B}\), which are analogous to the parameters \(\pi_{s1,A}\), \(\pi_{s1\_ B,A}\), and \(\pi_{s1\_ NA,A}\).

\newpage
\section*{Appendix 2 Derivation of the MLEs for Each Parameter in Table 2}

The likelihood contributions in Table 2 are based on the parameters
\[\Theta = \left\{ \phi,\phi_{A},\pi_{s1,A},\pi_{s1\_ B,A},\pi_{s1\_ NA,A},\pi_{s1,B},\pi_{s1\_ A,B},\pi_{s1\_ NA,B} \right\}\]
and we assume the additional parameters \(\psi\) and \(\xi_{A}\) are known from the design.

The vector of 17 cell counts in Table 2 is modeled as a multinomial
sample, i.e.,
\[\left( n_{1},\ n_{2},\ \cdots,n_{17} \right)\sim multinomial\left( N_{tot};\ p_{1},p_{2},\cdots,p_{17} \right)\]
Then the log-likelihood function with likelihood contributions provided in Table 2 is as follows,
\begin{align*}
    &l(\Theta;n_{1},\ n_{2},\ \cdots,n_{17}) \\
    & = \log\binom{N_{tot}}{n_{1},\ n_{2},\ \cdots,n_{17}} + n_{1}\log\left( p_{1} \right) + n_{2}\log\left( p_{2} \right) + \cdots n_{17}\log\left( p_{17} \right) \\
    & = \log\binom{N_{tot}}{n_{1},\ n_{2},\ \cdots,n_{17}} + n_{1}\left( \log\xi_{A} + \log\psi + \log\pi_{s1,A} + \log\phi_{A} + \log\phi \right)\\
    & + n_{2}\left[ \log\xi_{A} + \log\psi + \log{(1 - \pi_{s1,A})} + \log\phi_{A} + \log\phi \right] + \cdots\\
    &+ n_{17}\left[ \log{(1 - \psi)} + \log{(1 - \phi)} \right]
\end{align*}
   
\begin{itemize}
\item
  For \(\phi\), we take the derivative from
  \(l\left( \Theta;n_{1},\ n_{2},\ \cdots,n_{17} \right)\) and have the following equation:
  \[\frac{\partial}{\partial\phi}l\left( \Theta;n_{1},\ n_{2},\ \cdots,n_{17} \right) = \frac{\left( n_{1} + n_{2} + \cdots + n_{12} \right)}{\phi} - \frac{\left( n_{13} + n_{14} + \cdots + n_{17} \right)}{(1 - \phi)} = 0\ \]
Then we have \(\hat{\phi} = \frac{N_{1}}{N_{tot}}\), where
\(N_{1} = N_{tot} - (n_{13} + n_{14} + \cdots + n_{17})\), as well as
the variance estimator
\(\hat{V}(\hat{\phi}) = \left.\left\{ E\left\lbrack - \frac{\partial^{2}}{\partial\phi^{2}}l\left( \Theta;n_{1},\ n_{2},\ \cdots,n_{17} \right) \right\rbrack \right\}^{- 1} \right|_{\phi=\hat{\phi}} = \frac{\hat{\phi}\left( 1 - \hat{\phi} \right)}{N_{tot}}\).

\item 
For \(\phi_{A}\), we take the derivative regarding \(\phi_{A}\) and
  have the following equation:
\[\frac{\partial}{\partial\phi_{A}}l\left( \Theta;n_{1},\ n_{2},\ \cdots,n_{17} \right) = \frac{n_{1} + n_{2} + n_{3} + n_{4} + n_{11} + n_{12}}{\phi_{A}} - \frac{n_{5} + n_{6} + n_{7} + n_{8} + n_{9} + n_{10}}{\left( 1 - \phi_{A} \right)} = 0\ \]
Then we have \({\hat{\phi}}_{A} = \frac{N_{1,A}}{N_{1}}\), where
\(N_{1,A} = n_{1} + n_{2} + n_{3} + n_{4} + n_{11} + n_{12}\), as well as the variance estimator \(\hat{V} ({\hat{\phi}}_{A}) = \frac{{\hat{\phi}}_{A}\left( 1 - {\hat{\phi}}_{A} \right)}{N_{1}}\).

\item 
For \(\pi_{s1,A}\), we take the derivative regarding \(\pi_{s1,A}\),
  yielding the following equation:
\[\frac{\partial}{\partial\pi_{s1,A}}l\left( \Theta;n_{1},\ n_{2},\ \cdots,n_{17} \right) = \frac{n_{1} + n_{3}}{\pi_{s1,A}} - \frac{n_{2} + n_{4}}{\left( 1 - \pi_{s1,A} \right)} = 0\ \]
Then we have
\({\hat{\pi}}_{s1,A} = \frac{n_{1} + n_{3}}{n_{1} + n_{2} + n_{3} + n_{4}}\), as well as the variance estimator \(\hat{V}\left( {\hat{\pi}}_{s1,A} \right) = \frac{{\hat{\pi}}_{s1,A}\left( 1 - {\hat{\pi}}_{s1,A} \right)}{n_{1} + n_{2} + n_{3} + n_{4}}\).

\item 
For \(\pi_{s1\_ B,A}\), we take the derivative regarding
  \(\pi_{s1\_ B,A}\), yielding the following equation:
\[\frac{\partial}{\partial\pi_{s1\_ B,A}}l\left( \Theta;n_{1},\ n_{2},\ \cdots,n_{17} \right) = \frac{n_{5}}{\pi_{s1\_ B,A}} - \frac{n_{6}}{\left( 1 - \pi_{s1\_ B,A} \right)} = 0\ \]
Then we have \({\hat{\pi}}_{s1\_ B,A} = \frac{n_{5}}{n_{5} + n_{6}}\), as well as the variance estimator \(\hat{V}\left( {\hat{\pi}}_{s1\_ B,A} \right) = \frac{{\hat{\pi}}_{s1\_ B,A}\left( 1 - {\hat{\pi}}_{s1\_ B,A} \right)}{n_{5} + n_{6}}\).

\item 

For \(\pi_{s1\_ NA,A}\), we take the derivative regarding
  \(\pi_{s1\_ NA,A}\), yielding the following equation:
\[\frac{\partial}{\partial\pi_{s1\_ NA,A}}l\left( \Theta;n_{1},\ n_{2},\ \cdots,n_{17} \right) = \frac{n_{13}}{\pi_{s1\_ NA,A}} - \frac{n_{14}}{\left( 1 - \pi_{s1\_ NA,A} \right)} = 0\ \]
Then we have \({\hat{\pi}}_{s1\_ NA,A} = \frac{n_{13}}{n_{13} + n_{14}}\), as well as the variance estimator \(\hat{V}\left( {\hat{\pi}}_{s1\_ NA,A} \right) = \frac{{\hat{\pi}}_{s1\_ NA,A}\left( 1 - {\hat{\pi}}_{s1\_ NA,A} \right)}{n_{13} + n_{14}}\).

\end{itemize}

Similarly, we could evaluate the MLEs as well as the corresponding variances for the remaining two parameters in \(\Theta\), \(\pi_{s1\_ A,B}\) and \(\pi_{s1\_ NA,B}\) of Treatment \(B\) as well.

Of special note here is the fact that the covariances among the 8 closed-form MLEs above are all equal to zero under the multinomial model.

\newpage
\section*{Appendix 3 Derivation of Treatment Effects and Variance Estimators for Eqn.(2.3) and (2.4)}

Using the parameters defined in the previous appendix, we can derive the treatment effect estimators, \(\mu_{CRC}(A)\) and \(\mu_{CRC}(B)\). The following derivation for treatment \(A\) is based on Table S1.
\begin{align*}
    \mu_{CRC}(A) &= E[E(Y^A|\text{CRC scenario})] = E[Pr(Y^A = 1|\text{CRC scenario})] \\
    &= \sum_{i = 1}^{4} Pr(Y^{A} =1|\text{scenario $i$}) Pr(\text{scenario $i$}) \\
    &= \pi_{s1,A}\ \phi_{A}\phi\left\lbrack \xi_{A}\psi + (1 - \psi) \right\rbrack + \pi_{s1\_ B,A}\left\lbrack \left( 1 - \phi_{A} \right)\phi \right\rbrack\left\lbrack \xi_{A}\psi + (1 - \psi) \right\rbrack + \\
    &~~~ \pi_{s1\_ NA,A}(1 - \phi)\left\lbrack \xi_{A}\psi + (1 - \psi) \right\rbrack + \mu_{CRC}(A)(1 - \xi_{A})\ \psi
\end{align*}
With some algebra, the treatment effect estimator for \(A\) in eqn.(2.3) is derived as follows.
\[\mu_{CRC}(A) = \pi_{s1,A}\ \phi_{A}\phi + \pi_{s1\_ B,A}\ (1 - \phi_{A})\phi + \pi_{s1\_ NA,A}\ (1 - \phi).\]
Similarly, the treatment effect for \(B\) in eqn.(2.4) is written below
based on the proportions given in Table S2.
\[\mu_{CRC}(B) = \pi_{s1,B}\ (1 - \phi_{A})\phi + \pi_{s1\_ A,B}\ \phi_{A}\phi + \pi_{s1\_ NA,B}\ (1 - \phi).\]

\begin{table}[ht]
\centering
\caption*{Table S1: Scenarios for Evaluating the Treatment $A$ $^{a}$}
\renewcommand{\arraystretch}{1.3}
\setlength{\tabcolsep}{6pt}
\begin{tabular}{ccccccc}
\toprule
\multirow{3}{*}{Scenario} & \multirow{3}{*}{Stream 1} & \multirow{3}{*}{$P(S1)$} & \multirow{3}{*}{Stream 2} & \multirow{3}{*}{$P(S2)$} & Marginal  & \multirow{3}{*}{Proportion $^{b}$} \\
 & & & & & Treatment &
\\
 & & & & & Effect &
\\
\hline
\multirow{3}{*}{1} 
 & \multirow{2}{*}{Sampled and}  & \multirow{3}{*}{$\phi_A \phi$} & Sampled and  & \multirow{2}{*}{$\xi_A \psi$} & \multirow{3}{*}{$\pi_{S1,A}$} & \multirow{2}{*}{$\phi_A \phi [\xi_A \psi $} \\
 & \multirow{2}{*}{assigned A} & & randomized to A & & & \multirow{2}{*}{$+ (1-\psi)]$} 
 \\\cline{4-5}
 & & & Not sampled & $1-\psi$ & &  \\
\hline
\multirow{3}{*}{2} 
 & \multirow{2}{*}{Sampled and} & \multirow{3}{*}{$(1-\phi_A)\phi$} & Sampled and  & \multirow{2}{*}{$\xi_A \psi$} & \multirow{3}{*}{$\pi_{S1\_B,A}$} & \multirow{2}{*}{$(1-\phi_A)\phi$} \\
 & \multirow{2}{*}{assigned B}  & & randomized to A & & & \multirow{2}{*}{$[\xi_A \psi + (1-\psi)]$} \\\cline{4-5}
 & & & Not sampled & $1-\psi$ & & \\
\hline
\multirow{3}{*}{3} 
 & \multirow{3}{*}{Not sampled} & \multirow{3}{*}{$1-\phi$} & Sampled and & \multirow{2}{*}{$\xi_A \psi$} & \multirow{3}{*}{$\pi_{S1\_NA,A}$} & \multirow{2}{*}{$(1-\phi)[\xi_A \psi $} \\
 & & & randomized to A & & & \multirow{2}{*}{$+ (1-\psi)]$}\\\cline{4-5}
 & & & Not sampled & $1-\psi$ & & \\
\hline
\multirow{5}{*}{4} 
 & Sampled and & \multirow{2}{*}{$\phi_A \phi$} & \multirow{4}{*}{Sampled and} & \multirow{5}{*}{$(1-\xi_A)\psi$} & \multirow{5}{*}{$\mu_{CRC}(A)~^{c}$} & \multirow{5}{*}{$(1-\xi_A)\psi$} \\
 & assigned A & & \multirow{4}{*}{randomized to B} & & &  \\\cline{2-3}
 & Sampled and  & \multirow{2}{*}{$(1-\phi_A)\phi$} & & & & \\
 & assigned B &  & & & & \\\cline{2-3}
 & Not sampled & $1-\phi$ & & & & \\
\bottomrule
\end{tabular}

\vspace{0.6em}
\raggedright
\footnotesize \textit{$^{a}$ This table presents 4 scenarios and 9 sub-scenarios based on the observed types. For each stream, there are three possible situations: sampled and assigned/randomized to treatment A, sampled and assigned/randomized to treatment B, or not sampled in the data stream. This leads to 9 distinct combinations across two data streams. Some of these sub-scenarios can be grouped into 4 broader scenarios to facilitate the calculations. The overall treatment effect estimator is derived as a weighted average of each marginal treatment effect, based on the corresponding proportions. \\
$^{b}$ Under the Anchor-stream design, Stream 1 and Stream 2 are independent of each other. Therefore, the proportions for each scenario are calculated as Proportion (scenario $i$) = $\sum_{\text{scenario i}} P(S1)P(S2), \; i=1,2,3,4$. The sum of the proportions across the four scenarios equals 1. \\
$^{c}$ In the fourth scenario, Stream 2 randomizes individuals to a different treatment, rendering them unobserved due to the ``label-switching'' strategy. However, these individuals are a random sample from the target population, and the treatment effect is equivalent to the overall treatment effect.}
\end{table}

\begin{table}[ht]
\centering
\caption*{Table S2: Scenarios for Evaluating the Treatment $B$ $^{a}$}
\renewcommand{\arraystretch}{1.3}
\setlength{\tabcolsep}{6pt}
\begin{tabular}{ccccccc}
\toprule
\multirow{3}{*}{Scenario} & \multirow{3}{*}{Stream 1} & \multirow{3}{*}{$P(S1)$} & \multirow{3}{*}{Stream 2} & \multirow{3}{*}{$P(S2)$} & Marginal  & \multirow{3}{*}{Proportion $^{b}$} \\
 & & & & & Treatment &
\\
 & & & & & Effect &
\\
\hline
\multirow{3}{*}{1} 
 & \multirow{2}{*}{Sampled and}  & \multirow{3}{*}{$(1-\phi_A) \phi$} & Sampled and  & \multirow{2}{*}{$(1-\xi_A) \psi$} & \multirow{3}{*}{$\pi_{S1,B}$} & {$(1-\phi_A)\phi$} \\
 & \multirow{2}{*}{assigned $B$} & & randomized to $B$ & & & $[(1-\xi_A)\psi$ 
 \\\cline{4-5}
 & & & Not sampled & $1-\psi$ & & $+ (1-\psi)]$ \\
\hline
\multirow{3}{*}{2} 
 & \multirow{2}{*}{Sampled and} & \multirow{3}{*}{$\phi_A\phi$} & Sampled and  & \multirow{2}{*}{$(1-\xi_A) \psi$} & \multirow{3}{*}{$\pi_{S1\_A,B}$} & $\phi_A\phi$ \\
 & \multirow{2}{*}{assigned $A$}  & & randomized to $B$ & & &  $[(1-\xi_A)\psi$\\\cline{4-5}
 & & & Not sampled & $1-\psi$ & & $+ (1-\psi)]$\\
\hline
\multirow{3}{*}{3} 
 & \multirow{3}{*}{Not sampled} & \multirow{3}{*}{$1-\phi$} & Sampled and & \multirow{2}{*}{$(1-\xi_A) \psi$} & \multirow{3}{*}{$\pi_{S1\_NA,B}$} & $(1-\phi)$ \\
 & & & randomized to $B$ & & & $[(1-\xi_A)\psi$  \\\cline{4-5}
 & & & Not sampled & $1-\psi$ & & $+ (1-\psi)]$\\
\hline
\multirow{5}{*}{4} 
 & Sampled and & \multirow{2}{*}{$(1-\phi_A) \phi$} & \multirow{4}{*}{Sampled and} & \multirow{5}{*}{$\xi_A\psi$} & \multirow{5}{*}{$\mu_{CRC}(B)~^{c}$} & \multirow{5}{*}{$\xi_A\psi$} \\
 & assigned $B$ & & \multirow{4}{*}{randomized to $A$} & & &  \\\cline{2-3}
 & Sampled and  & \multirow{2}{*}{$\phi_A\phi$} & & & & \\
 & assigned $A$ &  & & & & \\\cline{2-3}
 & Not sampled & $1-\phi$ & & & & \\
\bottomrule
\end{tabular}

\vspace{0.6em}
\raggedright
\footnotesize \textit{$^{a}$ This table presents 4 scenarios and 9 sub-scenarios based on the observed types. For each stream, there are three possible situations: sampled and assigned/randomized to treatment $B$, sampled and assigned/randomized to treatment $A$, or not sampled in the data stream. This leads to 9 distinct combinations across two data streams. Some of these sub-scenarios can be grouped into 4 broader scenarios to facilitate the calculations. The overall treatment effect estimator is derived as a weighted average of each marginal treatment effect, based on the corresponding proportions. \\
$^{b}$ Under the Anchor-stream design, Stream 1 and Stream 2 are independent of each other. Therefore, the proportions for each scenario are calculated as Proportion (scenario $i$) = $\sum_{\text{scenario i}} P(S1)P(S2), \; i=1,2,3,4$. The sum of the proportions across the four scenarios equals 1. \\
$^{c}$ In the fourth scenario, Stream 2 randomizes individuals to a different treatment, rendering them unobserved due to the ``label-switching'' strategy. However, these individuals are a random sample from the target population, and the treatment effect is equivalent to the overall treatment effect.}
\end{table}

Since the derivation for all estimates of each parameter in \(\Theta\) has been provided above, the variance estimators for eqn.(2.3) and (2.4) can be readily evaluated using the multivariate delta method. Specifically, the variance can be computed as \(\hat{V} = g\hat{V}(\Theta)g^{T}\), leveraging the available closed-form variance-covariance matrix of \(\Theta\), i.e.,
\[
\hat{V}(\Theta) =
\begin{bmatrix}
\hat{V}(\hat{\phi}) & 0 & 0 & 0 & 0 & 0 & 0 & 0 \\
0 & \hat{V}(\hat{\phi}_A) & 0 & 0 & 0 & 0 & 0 & 0 \\
0 & 0 & \hat{V}(\hat{\pi}_{S1,A}) & 0 & 0 & 0 & 0 & 0 \\
0 & 0 & 0 & \hat{V}(\hat{\pi}_{S1,B,A}) & 0 & 0 & 0 & 0 \\
0 & 0 & 0 & 0 & \hat{V}(\hat{\pi}_{S1,NA,A}) & 0 & 0 & 0 \\
0 & 0 & 0 & 0 & 0 & \hat{V}(\hat{\pi}_{S1,B}) & 0 & 0 \\
0 & 0 & 0 & 0 & 0 & 0 & \hat{V}(\hat{\pi}_{S1,A,B}) & 0 \\
0 & 0 & 0 & 0 & 0 & 0 & 0 & \hat{V}(\hat{\pi}_{S1,NA,B})
\end{bmatrix}.
\]
For (2.3), we have
\[g = \begin{bmatrix}
{\hat{\pi}}_{s1,A}{\hat{\phi}}_{A} + {\hat{\pi}}_{s1\_ B,A}\ ( 1 - {\hat{\phi}}_{A}) - {\hat{\pi}}_{s1\_ NA,A} & \begin{matrix}
({\hat{\pi}}_{s1,A} - {\hat{\pi}}_{s1\_ B,A})\hat{\phi} & \begin{matrix}
{\hat{\phi}}_{A}\hat{\phi} & \begin{matrix}
(1 - {\hat{\phi}}_{A})\hat{\phi} & \begin{matrix}
1 - \hat{\phi} & 0 & 0 & 0 \\
\end{matrix} \\
\end{matrix} \\
\end{matrix} \\
\end{matrix} \\
\end{bmatrix}\]
and for (2.4) we have
\[g = \begin{bmatrix}
{\hat{\pi}}_{s1,B}(1 - {\hat{\phi}}_{A}) + {\hat{\pi}}_{s1\_ A,B}\ {\hat{\phi}}_{A} - {\hat{\pi}}_{s1\_ NA,B} & \begin{matrix}
( - {\hat{\pi}}_{s1,B} + {\hat{\pi}}_{s1\_ A,B})\hat{\phi} & \begin{matrix}
0 & \begin{matrix}
0 & 0 & \begin{matrix}
( 1 - {\hat{\phi}}_{A})\hat{\phi} & {\hat{\phi}}_{A}\hat{\phi} & 1 - \hat{\phi} \\
\end{matrix} \\
\end{matrix} \\
\end{matrix} \\
\end{matrix} \\
\end{bmatrix}.\]

%\newpage
\clearpage
\section*{Appendix 4 Table S3-S10 for Simulation Study}

\begin{table}[ht]
\centering
\caption*{Table S3: Simulation result to compare the estimation for treatment $A$ with $\mu_{\text{true}} = 0.68, \\ \, N_{\text{tot}}=500$}
%\label{tbl.3}
\renewcommand{\arraystretch}{1.3}
\setlength{\tabcolsep}{9pt}
\begin{tabular}{ccccccc}
\toprule
Setting & Estimation & $\hat{\mu}_{1}~^{a}$ & $\hat{\mu}_{RS}$ & $\hat{\mu}_{Chap}~^{b}$ & $\hat{\mu}_{\hat{\Psi}}~^{c}$ & $\hat{\mu}_{CRC}~^{c}$ \\
\midrule
\multirow{5}{*}{$p_{2}=5\%$}
 & mean   & 0.752 & 0.681 & 0.680 & 0.680 & 0.676 \\
 & SD     & 0.027 & 0.134 & 0.234 & 0.102 & 0.098 \\
 & Avg.SE & 0.027 & 0.127 & 0.183 & 0.091 & 0.097 \\
 & Width  & 0.107 & 0.498 & 0.511 & 0.356 (\textbf{0.336}) & 0.379 (\textbf{0.301}) \\
 & CI (\%) & 25.8 & 92.1 & 98.7 & 86.5 (\textbf{97.6}) & 85.9 (\textbf{97.0}) \\
\hline
\multirow{5}{*}{$p_{2}=10\%$}
 & mean   & 0.751 & 0.683 & 0.688 & 0.684 & 0.683 \\
 & SD     & 0.027 & 0.091 & 0.162 & 0.070 & 0.072 \\
 & Avg.SE & 0.028 & 0.091 & 0.136 & 0.067 & 0.066 \\
 & Width  & 0.108 & 0.356 & 0.449 & 0.263 (\textbf{0.256}) & 0.260 (\textbf{0.237}) \\
 & CI (\%) & 28.1 & 91.3 & 96.8 & 91.3 (\textbf{96.1}) & 89.1 (\textbf{94.9}) \\
\hline
\multirow{5}{*}{$p_{2}=20\%$}
 & mean   & 0.751 & 0.680 & 0.678 & 0.680 & 0.679 \\
 & SD     & 0.029 & 0.067 & 0.099 & 0.052 & 0.053 \\
 & Avg.SE & 0.028 & 0.065 & 0.093 & 0.050 & 0.050 \\
 & Width  & 0.111 & 0.256 & 0.356 & 0.195 (\textbf{0.203}) & 0.194 (\textbf{0.184}) \\
 & CI (\%) & 31.4 & 92.9 & 90.4 & 92.7 (\textbf{94.4}) & 92.3 (\textbf{93.6}) \\
\bottomrule
\end{tabular}

\vspace{0.6em}
\raggedright
\footnotesize \textit{a. the estimation result based on Stream 1 only is reported for $\hat{\mu}_{1}$ \\
b. the transformed logit CI (Sadinle, 2009) is reported for $\hat{\mu}_{Chap}$ \\
c. the proposed Bayesian Credible Interval (\textbf{bold}) is reported for $\hat{\mu}_{\hat{\Psi}}$ and $\hat{\mu}_{CRC}$}
\end{table}

\begin{table}[ht]
\centering
\caption*{Table S4: Simulation result to compare the estimation for treatment $B$ with $\mu_{\text{true}} = 0.54, \\ \, N_{\text{tot}}=500$}
%\label{tbl.3}
\renewcommand{\arraystretch}{1.3}
\setlength{\tabcolsep}{9pt}
\begin{tabular}{ccccccc}
\toprule
Setting & Estimation & $\hat{\mu}_{1}~^{a}$ & $\hat{\mu}_{RS}$ & $\hat{\mu}_{Chap}~^{b}$ & $\hat{\mu}_{\hat{\Psi}}~^{c}$ & $\hat{\mu}_{CRC}~^{c}$ \\
\midrule
\multirow{5}{*}{$p_{2}=5\%$}
 & mean   & 0.443 & 0.539 & 0.544 & 0.542 & 0.544 \\
 & SD     & 0.042 & 0.137 & 0.234 & 0.116 & 0.112 \\
 & Avg.SE & 0.041 & 0.133 & 0.225 & 0.109 & 0.109 \\
 & Width  & 0.159 & 0.520 & 0.728 & 0.427 (\textbf{0.394}) & 0.428 (\textbf{0.366}) \\
 & CI (\%) & 32.0 & 90.2 & 96.5 & 90.5 (\textbf{96.4}) & 88.0 (\textbf{95.8}) \\
\hline
\multirow{5}{*}{$p_{2}=10\%$}
 & mean   & 0.443 & 0.540 & 0.535 & 0.544 & 0.544 \\
 & SD     & 0.042 & 0.099 & 0.260 & 0.085 & 0.086 \\
 & Avg.SE & 0.041 & 0.098 & 0.221 & 0.081 & 0.079 \\
 & Width  & 0.162 & 0.383 & 0.666 & 0.318 (\textbf{0.303}) & 0.311 (\textbf{0.288}) \\
 & CI (\%) & 34.8 & 92.3 & 96.6 & 91.9 (\textbf{95.0}) & 89.8 (\textbf{94.2}) \\
\hline
\multirow{5}{*}{$p_{2}=20\%$}
 & mean   & 0.443 & 0.542 & 0.531 & 0.541 & 0.541 \\
 & SD     & 0.043 & 0.071 & 0.180 & 0.061 & 0.061 \\
 & Avg.SE & 0.042 & 0.070 & 0.158 & 0.059 & 0.058 \\
 & Width  & 0.166 & 0.273 & 0.552 & 0.231 (\textbf{0.231}) & 0.227 (\textbf{0.217}) \\
 & CI (\%) & 35.4 & 93.4 & 93.2 & 93.0 (\textbf{94.9}) & 92.8 (\textbf{94.4}) \\
\bottomrule
\end{tabular}

\vspace{0.6em}
\raggedright
\footnotesize \textit{a. the estimation result based on Stream 1 only is reported for $\hat{\mu}_{1}$ \\
b. the transformed logit CI (Sadinle, 2009) is reported for $\hat{\mu}_{Chap}$ \\
c. the proposed Bayesian Credible Interval (\textbf{bold}) is reported for $\hat{\mu}_{\hat{\Psi}}$ and $\hat{\mu}_{CRC}$}
\end{table}

\begin{table}[t]
\centering
\caption*{Table S5: Simulation result to compare the average treatment effect $^a$ with $ATE_{\text{true}} = 0.14, \\ N_{\text{tot}}=500$}
%\label{tbl.5}
\renewcommand{\arraystretch}{1.3}
\setlength{\tabcolsep}{9pt}
\begin{tabular}{ccccccc}
\toprule
Setting & Estimation & $ATE_{1}~^{b}$ & $ATE_{RS}$ & $ATE_{Chap}$ & $ATE_{\hat{\Psi}}~^{c}$ & $ATE_{CRC}~^{c}$ \\
\midrule
\multirow{5}{*}{$p_{2}=5\%$}
 & mean   & 0.310 & 0.142 & 0.136 & 0.139 & 0.132 \\
 & SD     & 0.051 & 0.193 & 0.330 & 0.155 & 0.150 \\
 & Avg.SE & 0.049 & 0.185 & 0.315 & 0.144 & 0.150 \\
 & Width  & 0.192 & 0.724 & 1.233 & 0.563 (\textbf{0.528}) & 0.588 (\textbf{0.480}) \\
 & CI (\%) & 8.0 & 92.8 & 98.3 & 90.9 (\textbf{96.7}) & 90.2 (\textbf{95.6}) \\
\hline
\multirow{5}{*}{$p_{2}=10\%$}
 & mean   & 0.309 & 0.143 & 0.152 & 0.140 & 0.139 \\
 & SD     & 0.050 & 0.132 & 0.305 & 0.108 & 0.110 \\
 & Avg.SE & 0.050 & 0.134 & 0.274 & 0.106 & 0.104 \\
 & Width  & 0.195 & 0.524 & 1.074 & 0.414 (\textbf{0.401}) & 0.408 (\textbf{0.375}) \\
 & CI (\%) & 8.2 & 94.3 & 96.3 & 93.2 (\textbf{95.7}) & 92.7 (\textbf{95.0}) \\
\hline
\multirow{5}{*}{$p_{2}=20\%$}
 & mean   & 0.308 & 0.138 & 0.147 & 0.138 & 0.138 \\
 & SD     & 0.052 & 0.097 & 0.207 & 0.078 & 0.079 \\
 & Avg.SE & 0.051 & 0.096 & 0.189 & 0.077 & 0.076 \\
 & Width  & 0.200 & 0.374 & 0.740 & 0.303 (\textbf{0.309}) & 0.299 (\textbf{0.285}) \\
 & CI (\%) & 10.0 & 93.9 & 94.7 & 94.1 (\textbf{94.9}) & 93.1 (\textbf{94.3}) \\
\bottomrule
\end{tabular}

\vspace{0.6em}
\raggedright
\footnotesize \textit{a. the average treatment effects (ATE) equals to $\hat{\mu}_A - \hat{\mu}_B$ \\
b. the estimation result based on Stream 1 only is reported for $\hat{\mu}_{1}$ \\
c. the proposed Bayesian Credible Interval (\textbf{bold}) is reported for ${ATE}_{\hat{\Psi}}$ and ${ATE}_{CRC}$}
\end{table}

\begin{table}[ht]
\centering
\caption*{Table S6: Simulation result to compare the estimation for treatment $A$ with $\mu_{\text{true}} = 0.68, \\ \, N_{\text{tot}}=5,000$}
%\label{tbl.3}
\renewcommand{\arraystretch}{1.3}
\setlength{\tabcolsep}{9pt}
\begin{tabular}{ccccccc}
\toprule
Setting & Estimation & $\hat{\mu}_{1}~^{a}$ & $\hat{\mu}_{RS}$ & $\hat{\mu}_{Chap}~^{b}$ & $\hat{\mu}_{\hat{\Psi}}~^{c}$ & $\hat{\mu}_{CRC}~^{c}$ \\
\midrule
\multirow{5}{*}{$p_{2}=5\%$}
 & mean   & 0.751 & 0.679 & 0.679 & 0.679 & 0.679 \\
 & SD     & 0.009 & 0.041 & 0.063 & 0.030 & 0.031 \\
 & Avg.SE & 0.009 & 0.042 & 0.062 & 0.031 & 0.031 \\
 & Width  & 0.034 & 0.163 & 0.252 & 0.120 (\textbf{0.120}) & 0.120 (\textbf{0.116}) \\
 & CI (\%) & 0.0 & 94.5 & 94.6 & 94.9 (\textbf{94.9}) & 94.1 (\textbf{94.7}) \\
\hline
\multirow{5}{*}{$p_{2}=10\%$}
 & mean   & 0.752 & 0.680 & 0.682 & 0.680 & 0.680 \\
 & SD     & 0.009 & 0.029 & 0.048 & 0.023 & 0.023 \\
 & Avg.SE & 0.009 & 0.029 & 0.044 & 0.022 & 0.022 \\
 & Width  & 0.034 & 0.115 & 0.170 & 0.086 (\textbf{0.089}) & 0.086 (\textbf{0.084}) \\
 & CI (\%) & 0.0 & 95.1 & 87.5 & 94.3 (\textbf{94.3}) & 94.1 (\textbf{94.4}) \\
\hline
\multirow{5}{*}{$p_{2}=20\%$}
 & mean   & 0.751 & 0.680 & 0.680 & 0.680 & 0.680 \\
 & SD     & 0.009 & 0.021 & 0.031 & 0.016 & 0.016 \\
 & Avg.SE & 0.009 & 0.021 & 0.030 & 0.016 & 0.016 \\
 & Width  & 0.035 & 0.082 & 0.118 & 0.062 (\textbf{0.067}) & 0.062 (\textbf{0.062}) \\
 & CI (\%) & 0.0 & 94.6 & 93.9 & 94.8 (\textbf{92.3}) & 94.7 (\textbf{94.6}) \\
\bottomrule
\end{tabular}

\vspace{0.6em}
\raggedright
\footnotesize \textit{a. the estimation result based on Stream 1 only is reported for $\hat{\mu}_{1}$ \\
b. the transformed logit CI (Sadinle, 2009) is reported for $\hat{\mu}_{Chap}$ \\
c. the proposed Bayesian Credible Interval (\textbf{bold}) is reported for $\hat{\mu}_{\hat{\Psi}}$ and $\hat{\mu}_{CRC}$}
\end{table}

\begin{table}[ht]
\centering
\caption*{Table S7: Simulation result to compare the estimation for treatment $B$ with $\mu_{\text{true}} = 0.54, \\ \, N_{\text{tot}}=5,000$}
%\label{tbl.3}
\renewcommand{\arraystretch}{1.3}
\setlength{\tabcolsep}{9pt}
\begin{tabular}{ccccccc}
\toprule
Setting & Estimation & $\hat{\mu}_{1}~^{a}$ & $\hat{\mu}_{RS}$ & $\hat{\mu}_{Chap}~^{b}$ & $\hat{\mu}_{\hat{\Psi}}~^{c}$ & $\hat{\mu}_{CRC}~^{c}$ \\
\midrule
\multirow{5}{*}{$p_{2}=5\%$}
 & mean   & 0.443 & 0.539 & 0.543 & 0.539 & 0.539 \\
 & SD     & 0.013 & 0.044 & 0.120 & 0.037 & 0.037 \\
 & Avg.SE & 0.013 & 0.044 & 0.112 & 0.037 & 0.036 \\
 & Width  & 0.051 & 0.174 & 0.439 & 0.144 (\textbf{0.143}) & 0.142 (\textbf{0.139}) \\
 & CI (\%) & 0.0 & 95.4 & 94.7 & 94.1 (\textbf{94.5}) & 94.1 (\textbf{94.3}) \\
\hline
\multirow{5}{*}{$p_{2}=10\%$}
 & mean   & 0.442 & 0.539 & 0.542 & 0.539 & 0.539 \\
 & SD     & 0.013 & 0.033 & 0.083 & 0.027 & 0.026 \\
 & Avg.SE & 0.013 & 0.032 & 0.079 & 0.026 & 0.026 \\
 & Width  & 0.051 & 0.123 & 0.309 & 0.103 (\textbf{0.104}) & 0.102 (\textbf{0.100}) \\
 & CI (\%) & 0.0 & 93.6 & 93.4 & 94.4 (\textbf{94.4}) & 94.3 (\textbf{94.5}) \\
\hline
\multirow{5}{*}{$p_{2}=20\%$}
 & mean   & 0.442 & 0.540 & 0.541 & 0.540 & 0.540 \\
 & SD     & 0.013 & 0.022 & 0.055 & 0.018 & 0.018 \\
 & Avg.SE & 0.013 & 0.022 & 0.054 & 0.019 & 0.019 \\
 & Width  & 0.053 & 0.087 & 0.211 & 0.074 (\textbf{0.076}) & 0.073 (\textbf{0.072}) \\
 & CI (\%) & 0.0 & 94.8 & 93.4 & 95.0 (\textbf{94.8}) & 95.0 (\textbf{95.2}) \\
\bottomrule
\end{tabular}

\vspace{0.6em}
\raggedright
\footnotesize \textit{a. the estimation result based on Stream 1 only is reported for $\hat{\mu}_{1}$ \\
b. the transformed logit CI (Sadinle, 2009) is reported for $\hat{\mu}_{Chap}$ \\
c. the proposed Bayesian Credible Interval (\textbf{bold}) is reported for $\hat{\mu}_{\hat{\Psi}}$ and $\hat{\mu}_{CRC}$}
\end{table}

\begin{table}[t]
\centering
\caption*{Table S8: Simulation result to compare the average treatment effect $^a$ with $ATE_{\text{true}} = 0.14, \\ N_{\text{tot}}=5,000$}
%\label{tbl.5}
\renewcommand{\arraystretch}{1.3}
\setlength{\tabcolsep}{9pt}
\begin{tabular}{ccccccc}
\toprule
Setting & Estimation & $ATE_{1}~^{b}$ & $ATE_{RS}$ & $ATE_{Chap}$ & $ATE_{\hat{\Psi}}~^{c}$ & $ATE_{CRC}~^{c}$ \\
\midrule
\multirow{5}{*}{$p_{2}=5\%$}
 & mean   & 0.309 & 0.140 & 0.136 & 0.140 & 0.140 \\
 & SD     & 0.015 & 0.061 & 0.135 & 0.049 & 0.048 \\
 & Avg.SE & 0.016 & 0.061 & 0.130 & 0.048 & 0.048 \\
 & Width  & 0.061 & 0.239 & 0.510 & 0.188 (\textbf{0.187}) & 0.186 (\textbf{0.181}) \\
 & CI (\%) & 0.0 & 95.1 & 95.8 & 94.5 (\textbf{95.0}) & 94.1 (\textbf{94.6}) \\
\hline
\multirow{5}{*}{$p_{2}=10\%$}
 & mean   & 0.309 & 0.141 & 0.140 & 0.141 & 0.141 \\
 & SD     & 0.016 & 0.043 & 0.096 & 0.035 & 0.035 \\
 & Avg.SE & 0.016 & 0.043 & 0.091 & 0.034 & 0.034 \\
 & Width  & 0.062 & 0.169 & 0.356 & 0.134 (\textbf{0.137}) & 0.133 (\textbf{0.131}) \\
 & CI (\%) & 0.0 & 95.2 & 95.2 & 94.5 (\textbf{94.5}) & 94.4 (\textbf{94.3}) \\
\hline
\multirow{5}{*}{$p_{2}=20\%$}
 & mean   & 0.309 & 0.140 & 0.140 & 0.140 & 0.140 \\
 & SD     & 0.016 & 0.031 & 0.064 & 0.024 & 0.024 \\
 & Avg.SE & 0.016 & 0.031 & 0.062 & 0.025 & 0.024 \\
 & Width  & 0.063 & 0.120 & 0.242 & 0.097 (\textbf{0.102}) & 0.096 (\textbf{0.095}) \\
 & CI (\%) & 0.0 & 95.2 & 94.5 & 95.0 (\textbf{93.4}) & 94.8 (\textbf{94.8}) \\
\bottomrule
\end{tabular}

\vspace{0.6em}
\raggedright
\footnotesize \textit{a. the average treatment effects (ATE) equals to $\hat{\mu}_A - \hat{\mu}_B$ \\
b. the estimation result based on Stream 1 only is reported for $\hat{\mu}_{1}$ \\
c. the proposed Bayesian Credible Interval (\textbf{bold}) is reported for ${ATE}_{\hat{\Psi}}$ and ${ATE}_{CRC}$}
\end{table}

\clearpage

\begin{table}[t]
\centering
\caption*{Table S9: Simulations Evaluating Mean Estimates for Continuous $X$ with $N_{\text{tot}}=1,000, \, p_{2}=5\%$}
%\label{tbl.6}
\renewcommand{\arraystretch}{1.3}
\setlength{\tabcolsep}{8pt}
\begin{tabular}{cccccccc}
\toprule
\multirow{2}{*}{Estimator} & {True} & \multirow{2}{*}{Methods} & \multirow{2}{*}{Mean} & \multirow{2}{*}{SD} & {Average} & {CI} & {Average} \\
 & mean & & & & SE $^a$ & Coverage\% & CI Width \\
\hline
\multirow{3}{*}{$\hat{\mu}_{A}~^{b}$} 
 & \multirow{3}{*}{5.020} 
 & Stream 1 only & 4.534 & 0.105 & -- & -- & -- \\
 &  & Stream 2 only & 5.024 & 0.602 & 0.589 & 93.1 & 2.290 \\
 &  & CRC & 5.015 & 0.505 & 0.484 & 92.1 & 1.890 \\
\hline
\multirow{3}{*}{$\hat{\mu}_{B}~^{b}$} 
 & \multirow{3}{*}{9.180} 
 & Stream 1 only & 9.409 & 0.185 & -- & -- & -- \\
 &  & Stream 2 only & 9.209 & 0.566 & 0.560 & 93.4 & 2.181 \\
 &  & CRC & 9.189 & 0.477 & 0.440 & 92.3 & 1.715 \\
\hline
\multirow{3}{*}{$\hat{\mu}^{AB}$} 
 & \multirow{3}{*}{-4.160} 
 & Stream 1 only & -4.875 & 0.214 & -- & -- & -- \\
 &  & Stream 2 only & -4.186 & 0.833 & 0.817 & 93.4 & 3.182 \\
 &  & CRC & -4.173 & 0.687 & 0.661 & 93.7 & 2.576 \\
\bottomrule
\end{tabular}

\vspace{0.6em}
\raggedright
\footnotesize \textit{a. SE for each estimator based on bootstrap with percentile CIs. \\
b. SE, CIs and their widths for the estimated mean not reported for the estimator based on Stream 1 only.}
\end{table}

\clearpage

\begin{table}[t]
\centering
\caption*{Table S10: Simulations Evaluating Mean Estimates for Continuous $X$ with $N_{\text{tot}}=1,000, \\ p_{2}=20\%$}
%\label{tbl.6}
\renewcommand{\arraystretch}{1.3}
\setlength{\tabcolsep}{8pt}
\begin{tabular}{cccccccc}
\toprule
\multirow{2}{*}{Estimator} & {True} & \multirow{2}{*}{Methods} & \multirow{2}{*}{Mean} & \multirow{2}{*}{SD} & {Average} & {CI} & {Average} \\
 & mean & & & & SE $^a$ & Coverage\% & CI Width \\
\hline
\multirow{3}{*}{$\hat{\mu}_{A}~^{b}$} 
 & \multirow{3}{*}{5.020} 
 & Stream 1 only & 4.524 & 0.114 & -- & -- & -- \\
 &  & Stream 2 only & 5.018 & 0.295 & 0.298 & 94.5 & 1.159 \\
 &  & CRC & 5.020 & 0.241 & 0.246 & 95.7 & 0.958 \\
\hline
\multirow{3}{*}{$\hat{\mu}_{B}~^{b}$} 
 & \multirow{3}{*}{9.180} 
 & Stream 1 only & 9.414 & 0.188 & -- & -- & -- \\
 &  & Stream 2 only & 9.166 & 0.280 & 0.280 & 95.0 & 1.090 \\
 &  & CRC & 9.171 & 0.223 & 0.227 & 94.4 & 0.882 \\
\hline
\multirow{3}{*}{$\hat{\mu}^{AB}$} 
 & \multirow{3}{*}{-4.160} 
 & Stream 1 only & -4.890 & 0.220 & -- & -- & -- \\
 &  & Stream 2 only & -4.149 & 0.402 & 0.410 & 96.1 & 1.597 \\
 &  & CRC & -4.151 & 0.321 & 0.336 & 96.3 & 1.301 \\
\bottomrule
\end{tabular}

\vspace{0.6em}
\raggedright
\footnotesize \textit{a. SE for each estimator based on bootstrap with percentile CIs. \\
b. SE, CIs and their widths for the estimated mean not reported for the estimator based on Stream 1 only.}
\end{table}

\clearpage
\section*{Appendix 5 Reliability of the Illustrative Example}

In Section 4, an illustrative data example is provided to demonstrate use of the proposed methods to compare vaccine response rates. Recognizing that a single synthetic data example cannot adequately reflect the generalizability of the proposed estimators, we generated 1,000 synthetic datasets under the same conditions and following the same procedure outlined in the text. That is, we kept the first 1,000 datasets that directly mimicked Stream 2 in the example in the sense that exactly 85 seropositive responses for vaccine \(A\) and 71 for vaccine \(B\) were obtained. The robustness and precision of the proposed CRC estimator was then evaluated against the random sampling-based estimator, with results summarized in Table S11.

\begin{table}[ht]
\centering
\caption*{Table S11: Seropositivity Estimates and Comparison for the Synthetic Population in Tunisia $^{a}$}%\label{tbl.8}
\renewcommand{\arraystretch}{1.3}
\setlength{\tabcolsep}{10pt}
\begin{tabular}{cccccc}
\hline
\multirow{2}{*}{Vaccine} & \multirow{2}{*}{Estimator} & \multirow{2}{*}{Mean} & \multirow{2}{*}{Avg.SE} & \multirow{2}{*}{Avg.Width $^{b}$} & Overall  \\
& & & & & Proportion $^c$ \\
\hline
\multirow{2}{*}{Vaccine A}
 & $\hat{\mu}_{RS,A}$   & 98.8\% & 0.0116 & 0.034 $^{d}$ & \multirow{2}{*}{73.8\%} \\
 & $\hat{\mu}_{CRC,A}$  & 98.1\% & 0.0085 & 0.024 & \\
\hline
\multirow{2}{*}{Vaccine B}
 & $\hat{\mu}_{RS,B}$   & 85.5\% & 0.0386 & 0.151 & \multirow{2}{*}{100\%} \\
 & $\hat{\mu}_{CRC,B}$  & 86.4\% & 0.0325 & 0.046 & \\
\hline
\multirow{2}{*}{Difference $^{e}$}
 & $ATE_{RS}$  & 13.3\% & 0.0403 & 0.158 & \multirow{2}{*}{98.5\%} \\
 & $ATE_{CRC}$ & 11.8\% & 0.0337 & 0.132 & \\
\hline
\end{tabular}

\vspace{0.6em}
\raggedright
\footnotesize \textit{a. The analysis is based on Tunisian trial data reported in Ben Ahmed and others (2022). The results for the RS-based estimators remain the same across replicates, as the seropositive responses were fixed to mimic Stream 2 data reported in Ben Ahmed and others (2022). \\
b. The widths reported for $\hat{\mu}_{RS,A}$ and $\hat{\mu}_{RS,B}$ are based on the Wald-type 95\% CI; the widths reported for $\hat{\mu}_{CRC,A}$ and $\hat{\mu}_{CRC,B}$ are based on the proposed Bayesian Credible Intervals in Section 2.4. \\
c. The overall proportion indicates the probability that the CI width for the RS approach exceeds that of the CRC approach across all replicates regarding Vaccine A, B, or their difference (ATE). \\
d. The upper limit of the CI for $\hat{\mu}_{RS,A}$ is capped at 100\%, with an average width of 0.045 when disregarding the cap. \\
e. The difference of two vaccines equals to $\hat{\mu}_{RS,A} - \hat{\mu}_{RS,B}$ or $\hat{\mu}_{CRC,A} - \hat{\mu}_{CRC,B}$, which matches the definition of ATE introduced in previous sections.}
\end{table}

The mean estimates across simulations, average standard errors (SE), and average interval widths are presented in Table S11, demonstrating that the CRC estimators exhibit less bias compared to the RS estimator. To assess generalizability, width differences across all replicates are summarized in the last column. In most cases, the CRC estimators yield narrower confidence or credible intervals as proposed in the article. Note that for Vaccine \(A\), the overall proportion of intervals favoring the CRC approach is perhaps artificially low, due to capping the upper limit of the confidence interval for \({\hat{\mu}}_{RS,A}\) at 100\%. This proportion increased to 99.1\% across 1,000 simulations when the natural limit was disregarded. In contrast, the credible intervals from the CRC approach consistently respect the natural range of [0, 1], eliminating such concerns entirely.